\documentclass[twocolumn]{aastex63}
\shorttitle{Irrad BDs}
\shortauthors{Lothringer \& Casewell}

\usepackage{amsmath}
\usepackage{xspace}
\usepackage{float}
\usepackage{array}

\providecommand{\adsurl}[1]{\href{#1}{ADS}}
\def\aap{{A\&A}}		
\def\apj{{ApJ}}			
\def\apjl{{ApJ}}		
\def\apjs{{ApJS}}		
\def\pasp{{PASP}}		

\def\mnras{{MNRAS}}
\def\nat{{Nature}}
\def\aj{{Astronomical Journal}}

\def\rsun{{\rm\,R_\odot}}

\newcommand{\microns}{$\mu$m}

\newcommand{\logg}{$\log(g)$}

\newcommand{\mjup}{${\rm M}_{\rm Jup}$}

\begin{document}
	\title{Atmosphere Models of Brown Dwarfs Irradiated by White Dwarfs: Analogues for Hot and Ultra-Hot Jupiters}
	\author[0000-0003-3667-8633]{Joshua D. Lothringer}
	\affiliation{Department of Physics and Astronomy, Johns Hopkins University, Baltimore, MD 21210 USA}

	\author[0000-0003-2478-0120]{Sarah L. Casewell}
	\affiliation{School of Physics and Astronomy, University of Leicester, University Road, Leicester LE1 7RH, UK}

	\vspace{0.5\baselineskip}
	\date{\today}
	\email{jlothrin@lpl.arizona.edu}

	\begin{abstract}
		
	Irradiated brown dwarfs (BDs) provide natural laboratories to test our understanding of substellar and irradiated atmospheres. A handful of short-period BDs around white dwarfs (WDs) have been observed, but the uniquely intense UV-dominated irradiation presents a modeling challenge. Here, we present the first fully self-consistent 1D atmosphere models that take into account the UV irradiation's effect on the object's temperature structure. We explore two BD-WD systems, namely WD-0137-349 and EPIC-212235321. WD-0137-349B has an equilibrium temperature that would place it in the transition between hot and ultra-hot Jupiters, while EPIC-212235321B has an equilibrium temperature higher than all ultra-hot Jupiters except KELT-9b. We explore some peculiar aspects of irradiated BD atmospheres and show that existing photometry can be well-fit with our models. Additionally, the detections of atomic emission lines from these BDs can be explained by a strong irradiation-induced temperature inversion, similar to inversions recently explored in ultra-hot Jupiters. Our models of WD-0137-349B can reproduce the observed equivalent width of many but not all of these atomic lines. We use the observed photometry of these objects to retrieve the temperature structure using the PHOENIX ExoplaneT Retrieval Algorithm (PETRA) and demonstrate that the structures are consistent with our models, albeit somewhat cooler at low pressures. We then discuss the similarities and differences between this class of irradiated brown dwarf and the lower-mass ultra-hot Jupiters. Lastly, we describe the behavior of irradiated BDs in color-magnitude space to show the difficulty in classifying irradiated BDs using otherwise well-tested methods for isolated objects. 
	\end{abstract}
	

	\section{Introduction} \label{section:intro}
	
Highly irradiated brown dwarfs (BDs) are rare. Only about 0.24$\pm$0.04\% of solar-type stars have a brown dwarf companion with a period less than 80 days, compared to $>${1.7}$\pm$0.5\% that have a brown dwarf companion with a period less than 1000 days \citep{kiefer:2019}. While short-period brown dwarf companions are not a common outcome of star formation and evolution, such substellar objects provide a pivotal comparison to both other irradiated objects and isolated brown dwarfs. Irradiated brown dwarfs experience both high external irradiation and interior heat flux, comparable only to young hot Jupiters \citep{showman:2016nandv}. This unique regime can therefore be probed by such objects and, currently, the interplay between high levels of irradiation and interior heat is not well understood or observed. Thus, these highly irradiated brown dwarfs provide essential tests to models of brown dwarf evolution and atmospheric physics.

\subsection{White Dwarf-Brown Dwarf Pairs}

A physically interesting and observationally favorable class of irradiated brown dwarfs are those found in ultra-short-periods around white dwarfs (WDs). These objects typically have periods on the order of hours around T$_{eff}>${10$^4$~K} white dwarfs and can have equilibrium temperatures over 3000~K. Some of these systems are interacting (i.e., mass transfer is taking place) while others seem to be non-interacting. The short periods of these systems are thought to come from common envelope (CE) evolution in which the secondary orbited within the primary's envelope during the primary's\added{ post-main sequence} giant phase \citep{iben:1993a,iben:1993b}. The interaction between the secondary and the envelope shrunk the orbit of the secondary while shedding the primary's envelope, resulting in the short-period BD-WD system seen today \citep{rasio:1996,politano:2004}.

Only a handful of these BD-WD systems are known today, but several are benchmark objects for understanding highly irradiated atmospheres. About 10 non-interacting post-common-envelope BD-WD binaries are known with several more interacting systems \citep{casewell:2018b}.

\subsection{Irradiated Brown Dwarfs as High-Mass Hot Jupiters}

Short-period, highly-irradiated brown dwarfs are known to reach equilibrium temperatures up to 3000~K \citep{rappaport:2017,casewell:2018}. This means that these ultra-hot brown dwarfs are the only \added{known }gaseous substellar objects with equilibrium temperatures between the hottest known hot Jupiter, KELT-9b ($T_\textrm{{eq}}$=4050 K, \cite{gaudi:2017}), and the next hottest hot Jupiter, WASP-33b ($T_\textrm{{eq}}$ = 2800 K, \cite{colliercameron:2013}). These ultra-hot brown dwarfs thus provide crucial insight into the behavior of highly irradiated atmospheres.

Highly-irradiated brown dwarfs around white dwarfs also offer unique insight into the effects of different irradiation spectra on an atmosphere. White dwarfs with known short-period brown dwarf companions vary between T$_{eff}$= 7000 - 37,000~K. Meanwhile, KELT-9 is the hottest known exoplanet host with T$_{eff}$= 10,170 K. For a brown dwarf like EPIC212235321B, which orbits a T$_{eff}$= 25,000 K white dwarf, 91\% of the radiation it absorbs from its host is in the UV ($<${4000} \AA). How a substellar atmosphere responds to such intense short-wavelength irradiation has been neither widely observed, nor modeled in detail. Such systems provide opportunities to understand how substellar atmospheres behave around early-type host stars.

Brown dwarf companions to white dwarfs also have some observational advantages compared to ultra-hot Jupiters, which themselves are some of the best targets for observation because of their short periods, bright daysides, and inflated radii. While the total system brightness of BD-WD pairs is low (usually K=15-20 mag for known systems), the actual number of photons measured from the brown dwarfs is comparable, and in some instances greater than, the number of photons measured from hot Jupiter systems. This, combined with the fact that there is less photon noise from the small white dwarf, means that both the dayside and nightside of these objects can be measured well. Lastly, the ultra-short period of these brown dwarfs means that several full phase curves can be observed from the ground in a single night, something currently impossible for hot Jupiters.

\subsection{Exoplanets Around White Dwarfs}

\added{The recent discovery of a short-period giant planet or very low-mass brown dwarf around the nearby WD 1856+534 highlights the connection between BDs in WD systems and exoplanets \citep{vanderburg:2020}. Constraints on the companions mass from its lack of thermal emission and age place it likely below 11 \mjup{}, demonstrating that low mass objects can survive the host stars post-main sequence evolution. The small radius of the white dwarfs can result in very large, often total eclipse of the star by the planet. This provides a unique opportunity to potentially characterize even terrestrial planets \citep{kozakis:2020,kaltenegger:2020}.}

\subsection{Models of White Dwarf Companions}

While the UV irradiation of a brown dwarf by a white dwarf has not before been self-consistently modeled, some attention has been given to M-dwarfs around white dwarfs. \cite{brett:1993} and \cite{barman:2004} both modeled such atmospheres, including the systems GD 245, NN Ser, AA Dor, and UU Sge. Both works found that temperature inversions of thousands of Kelvin can form on the irradiated hemisphere of the companions. Such models focused exclusively on irradiated stellar atmospheres, with the fundamental difference in modeling irradiated stellar versus substellar atmosphere being the difference in interior heat (i.e., a larger proportion of an irradiated brown dwarf's $T_{\textrm{eff}}$ will be determined by the irradiation), as well as the minimum temperatures in the atmosphere.

The models we describe in this work are similar to these models but have benefited from years of improvements in the development of molecular line lists, the discovery and observation of lower-mass white dwarf companions, and experience in modeling the highly irradiated atmospheres of ultra-hot Jupiters \citep[e.g.,][]{lothringer:2018b,lothringer:2019}.

\subsection{Outline}

In this work, we calculate atmosphere models of brown dwarfs by white dwarfs in order to understand the behavior of such highly irradiated atmospheres, compare model spectra to observations of several such systems, and place them into context with their lower-mass irradiated cousins, the hot Jupiters. We describe our model in more detail in Section~\ref{methods}. We then compare models of two highly irradiated brown dwarfs, WD-0137-349B\added{ (hereafter, WD-0137B)} and EPIC212235321B\added{ (hereafter, EPIC2122B)}, to observations in Section~\ref{results}, including a retrieval analysis. We then discuss these results in the broader context of substellar atmospheres in Section~\ref{discuss} and conclude in Section~\ref{conclusion}.

\section{Methods}\label{methods}

The PHOENIX atmosphere model provides an ideal platform to model highly irradiated atmospheres. PHOENIX is capable of modeling a wide range of objects from cool planets to hot stars thanks in large part to a comprehensive opacity database with opacities from the UV to the FIR. While PHOENIX has been widely used to model isolated objects \citep{hauschildt:1999,allard:2014,barman:2015}, it has also been applied to highly irradiated stars \citep{barman:2004} and exoplanets \citep{barman:2001,barman:2002,lothringer:2018b,lothringer:2019,lothringer:2020b}.

\added{PHOENIX works by first calculating chemical equilibrium using solar metallicity elemental abundances from \cite{asplund:2005:abund} for a given 64-layer pressure-temperature structure on an log-spaced optical depth grid that extends from $\tau = 10^{-8}$ to 10$^{2.5}$. This generally corresponds to pressures of about $10^{-10}$ to $10^{2}$ bars. Here, we assume the atmosphere is in local thermodynamic equilibrium (LTE), i.e., that collisions dominate the determination of the atomic and molecular level populations. Future studies may explore the potential for the high levels of irradiation to bring the atmosphere out of LTE.}

\added{The opacity is then calculated at each point on the wavelength grid, which extends from 10 to 10$^6$~\AA{} using direct-opacity sampling (dOS) \citep{schweitzer:2000}. The short-wavelength portion of the spectrum is crucial to determining the correct temperature structure of the BD atmosphere due to the high UV and optical irradiation from the white dwarf host star. We include opacity from 130 different molecular species plus atomic opacity from elements up to uranium. Many continuous opacity sources, from collision-induced-absorption to H$^{-}$ opacity, are also included. In Section~\ref{results}, most models are cloud-free as adding clouds did not significantly improve the fit to observations, but we do include equilibrium clouds in Section~\ref{discuss} when we investigate the behavior of irradiated BDs in color-magnitude space. For these cloudy models, we use a mean particle radius of 1~\microns{} and a log-normal size distribution.}

Radiative transfer is then calculated including the irradiation by the white dwarf host star using the closest matching white dwarf spectrum from \cite{koester:2010}. The white dwarf spectra extend from 900 to 30,000~\AA{}. From this, the vertical flux is calculated and the pressure-temperature structure is modified accordingly to bring the model closer to radiative equilibrium using a modified Uns\"old-Lucy method \citep{hauschildt:2003}. This whole processes is then repeated iteratively until the maximum temperature correction for any layer is less than 0.5~K.

We use the parameter $f$ to define the redistribution of heat across the planet from the dayside to the nightside. In our 1D model, this determines the fraction of irradiation that goes into heating the part of atmosphere we are modeling. We first define the quantity $T_{\textrm{irr}}$, which is the effective temperature of the BD if there no internal heat. $T_{\textrm{irr}}$ is determined by the stellar, orbital, and atmospheric properties thusly:

\begin{equation}\label{eq1}
		T_{\textrm{irr,BD}} = (f*(1-A_{\textrm{BD}}))^{1/4}*T_{\textrm{eff,WD}} \sqrt{R_{\textrm{WD}}/a}
\end{equation}
\noindent where $f$ is a heat redistribution factor, $A$ is the Bond albedo, and $a$ is the orbital distance. The final effective temperature of the BD, T$_{\textrm{eff,BD}}$, is then

\begin{equation}
	T_{\textrm{eff,BD}} = \sqrt[4]{T_{\textrm{int,BD}}^4+T_{\textrm{irr,BD}}^4}.
\end{equation}
\noindent where $T_{\textrm{int,BD}}$ is the internal temperature of the BD.

For each object, we calculated models corresponding to full or dayside heat redistribution corresponding to $f=0.25$ and 0.5, respectively. For WD-0137B, we found it necessary to include models cooler than the full heat redistribution models so we include a model that are about 250~K cooler. Furthermore, we assume an albedo, $A$ of zero, but this is degenerate with the heat redistribution here. We describe possible implications of this in Section~\ref{discuss}. Table~\ref{tab:models} shows a summary of the computed models. Properties of the primary and secondary objects, e.g., $log$(g) and the orbital parameters, are taken from the literature for each object \citep{maxted:2006,casewell:2015,casewell:2018} and are listed in Table~\ref{table:properties}. We choose to limit the number of self-consistent models computed since we also use a retrieval, described below, to provide a better fit to the data.

For WD-0137B, we chose an internal temperature of 1400~K, consistent with the BD's inferred mass, radius, and previously inferred spectral type \citep{maxted:2006,burleigh:2006,casewell:2015}. For EPIC2122B, we modelled two internal temperatures of 1000 and 2000~K. An internal temperature of 2000~K is consistent with previous spectral typing from \citep{casewell:2018}, while a temperature of 1000~K is more comparable to expectations of similarly irradiated exoplanets \citep{thorngren:2019}.

 \begin{table*}[t] 
	\centering  
	\caption{Model Summary for Spectrum Fits}
	\label{tab:models} 
	\begin{tabular}{cccccccc}
		\hline
		Target & Model Name & Redistribution $f$ & $T_{\textrm{int}}$ (K) & $T_{\textrm{irr}}$ (K) & $T_{\textrm{eff}}$ (K) &  Clouds & $\chi^2$ per point  \\
		\hline 
		\hline
		WD-0137-347B & 1A & 0.15 & 1,400 & 1,660 & 1,840 & No &  2.7  \\
		 $T_{\textrm{eq}}$ = 1,990~K   & 1B & 0.15 & 1,400 & 1,660 & 1,840 & Yes  &  2.8 \\
		    & 1C & 0.25 & 1,400 & 1,990 & 2,100 & No  &  7.36\\
		    & 1D & 0.25 & 1,400 & 1,990 & 2,100 & Yes   & 18.6 \\
    		& 1E & 0.5 & 1,400 & 2,360 & 2,430 & No  &  27.3\\ 
				 & Retrieval  & --- & 684$^{+481}_{-217}$ & 1134$^{+298}_{-338}$ & $1253^{+300}_{-291}$ & No & 1.5\\
		\hline 
		EPIC212235321B & 2A & 0.25 & 1,000 & 3,435 &3,450 & No &  41.4 \\
		$T_{\textrm{eq}}$ = 3,435~K & 2B & 0.25 & 2,000 & 3,435 &3,525 & No & 26.1  \\
		& 2C & 0.5 & 1,000  & 4,040 &4,050 & No  & 258 \\
		 & 2D &0.5 & 2,000 & 4,040 &4,125 & No  & 271 \\
		 & Retrieval & --- & $475^{+322}_{-216}$ & $3425^{+376}_{-411}$ & $3430^{+375}_{-407}$ & No & 18.6 \\
		\hline

	\end{tabular}
	\tablecomments{A redistribution of $f$=0.25 corresponds to planet-wide heat redistrbution, while $f$=0.5 corresponds to dayside-only heat redistribution. Also note that the value of $T_{\textrm{int}}$ in the retrieval can be somewhat degenerate with the shape of the temperature structure in the parameterization used (see Section~\ref{methods}).}
\end{table*} 

While photoionization cross-sections are included in the opacity calculation, photoionization and photochemistry are not directly included in the determination of the chemical abundances. While the hot atmospheres of the objects we study likely have short chemical timescales \citep{kitzmann:2018}, the extreme UV flux these brown dwarfs experience likely has some effect on the chemistry. Indeed, photochemical modeling in \citep{lee:2020} indicate significant photo-ionization and dissociation at pressures below 0.1 mbar.
\subsection{Retrievals with PETRA}

In addition to self-consistent forward models, we also fit the photometry using the PHOENIX ExoplaneT Retrieval Analysis, or PETRA \citep{lothringer:2020a}. PETRA is a retrieval framework built around PHOENIX and allows us to vary atmospheric parameters to match the data. By exploring parameter space, retrievals can provide statistically robust uncertainties on atmospheric properties. By combining our self-consistent atmosphere models with a retrieval within the same modeling foundation, we can compare both our theoretical expectations for what the atmosphere should look like with what the data actually imply about the atmosphere.

\added{In a retrieval framework, spectra are calculated for many different combinations of atmospheric parameters and then compared to observations. Statistical methods are used to efficiently explore parameter space and generate posterior distributions for the modelled parameters. PETRA uses a custom-built Differential Evolution Markov Chain framework \citep{terbraak:2006,terbraak:2008}, a type of Markov Chain Monte Carlo (MCMC). Because many iterations need to be calculated, assumptions, like uniform vertical chemical abundances, and parameterizations, like those used to define the temperature structure, are often used to reduce the dimensionality of the parameter space explored.}

In this work, since we rely solely on comparing to a small number of broadband photometry observations, we chose to only fit the temperature structure, while keeping parameters like the metallicity and surface gravity fixed. Specifically, the atomic and molecular abundances in the retrievals are in chemical equilibrium and assume solar metallicity and elemental abundances from \cite{asplund:2005}. Additionally, the retrieval models are cloud-free.

We use the temperature structure parameterization from \cite{parmentier:2014} while including the internal temperature as an additional free parameter. In total, six parameters are used to describe the temperature structure, allowing for enough flexibility to fit a structure with both an internal adiabat and a temperature inversion, while requiring a reasonable number of free parameters. Note that the temperature structures in the retrieval are not necessarily in radiative equilibrium; rather, the retrieval allows the data to drive the atmospheric properties. \added{The internal temperature is somewhat degenerate with the shape of the transit spectrum, but we include it for increased flexibility in fitting the data.}

\added{Prior information can also be added into the Bayesian statistical framework to constrain the atmospheric estimated parameters. We generally place wide, uniform priors on the estimated parameters to keep the mean temperature of the atmosphere above 500~K and below 10,000~K. A prior is also placed to keep the internal temperature between 100~K and 4000~K. In the end, these priors do not affect the outcome of the retrievals since the retrieved parameters are well within the wide range allowed by the priors.}

Because of the simplicity of the retrievals (i.e., low number of free parameters), convergence was achieved for both the WD-0137-349B and the EPIC212235321B retrievals in about 3,000 iterations on 10 chains, confirmed with a Gelman-Rubin statistic of $<1.01$ \citep{gelman:1992} and by visual inspection of the chains, throwing away the first 500 iterations as initialization burn-in.

\section{Results}\label{results}

 \begin{table*}[t] 
	\centering  
	\caption{Model Parameters}
	\label{table:properties} 
	\begin{tabular}{cccccccc}
		\hline
		Target & Separation ($\rsun$) & T$_{\textrm{eff,WD}}$ (K) & R$_{\textrm{WD}}$ ($\rsun$) & \logg$_{\textrm{WD}}$ & T$_{\textrm{int,BD}}$ (K) & R$_{\textrm{BD}}$ ($\rsun$) &  \logg$_{\textrm{BD}}$ \\
		\hline 
		\hline
		WD-0137-347B & 0.65 & 16,500 & 0.0186   & 7.5 & 1,400 & 0.1   & 5.1  \\ 
		\hline 
		EPIC212235321B & 0.44 & 24,900 & 0.017   & 7.65 & 1,000-2,000 & 0.0973   & 5.2  \\ 
		\hline

	\end{tabular}

\end{table*}

\subsection{WD 0137-349B}\label{sec:wd01347}

\begin{figure*}[ht!]
	\centering
	\includegraphics[width=3.3in]{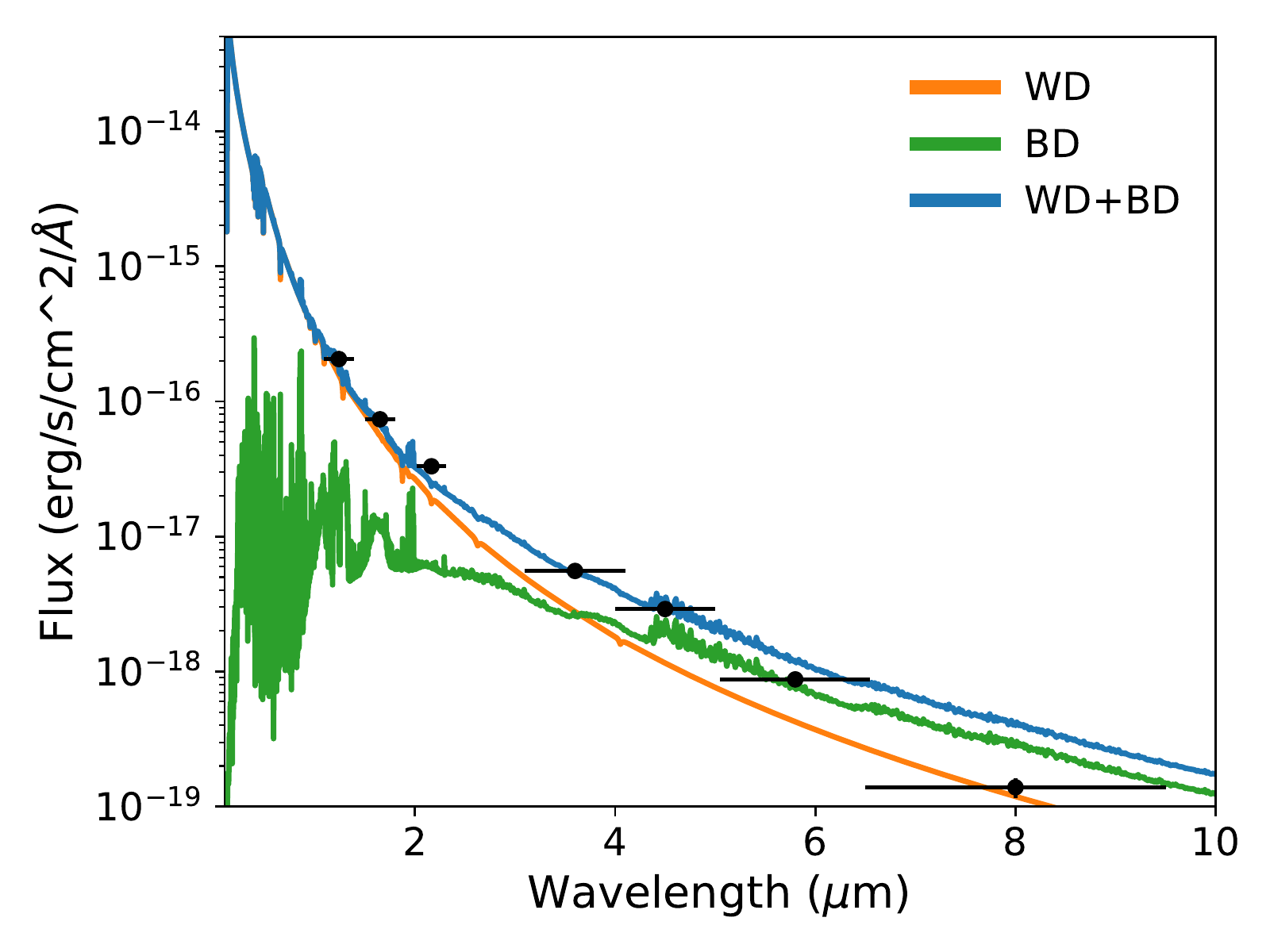} 
	\includegraphics[width=3.3in]{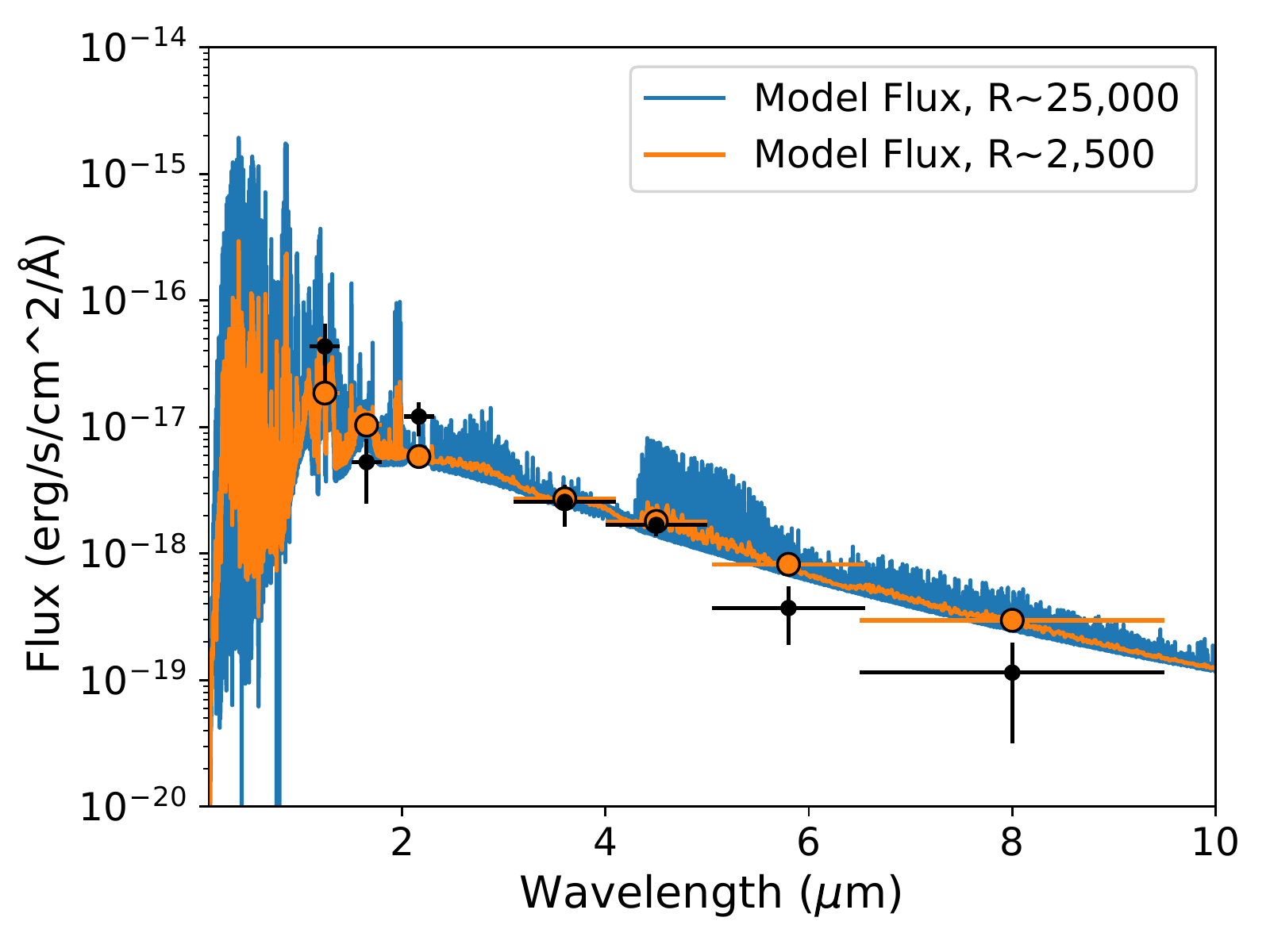}
	\caption{\small The best fitting spectra of the WD~0137-349 system\added{ from Model~1A}. Left shows the WD spectrum (orange), BD spectrum (green), and their combined spectrum (blue) compared to photometry from \citep{casewell:2015}. Right shows the BD spectrum at high (R$\sim$25,000 at 2.5~\microns{}, blue) and medium (R$\sim$2,500, orange) spectral resolution compared to the same photometry but with the WD contribution subtracted. \label{fig:wd0137_spec}}
\end{figure*}

\begin{figure*}[ht!]
	\centering
	\includegraphics[width=5.5in]{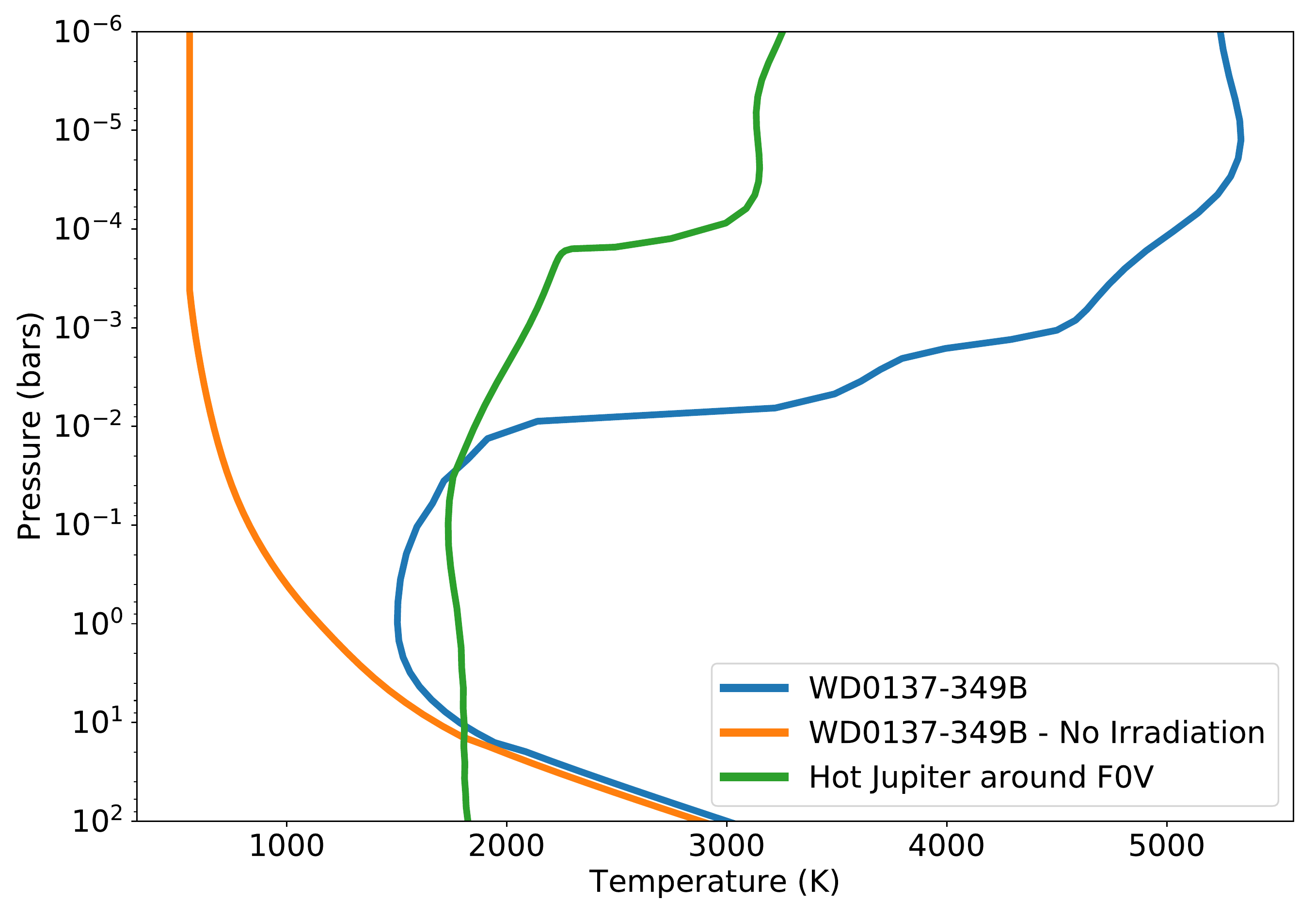}
	\caption{\small Temperature structures from the WD-0137-349B atmosphere model comparing the best-fitting self-consistent case (Model 1A), an identical atmosphere without irradiation, and a hot Jupiter of similar effective temperature (T$_{eff}\sim$1900~K, T$_{int}\sim$125~K) irradiated by a T$_{\textrm{eff}}$=7200~K F0V star. Note the uniquely strong inversion \textit{and} high internal temperature of WD-0137-349B. \label{fig:wd0137_TP}}
\end{figure*}

\begin{figure}[ht!]
	\centering
	\includegraphics[width=3.5in]{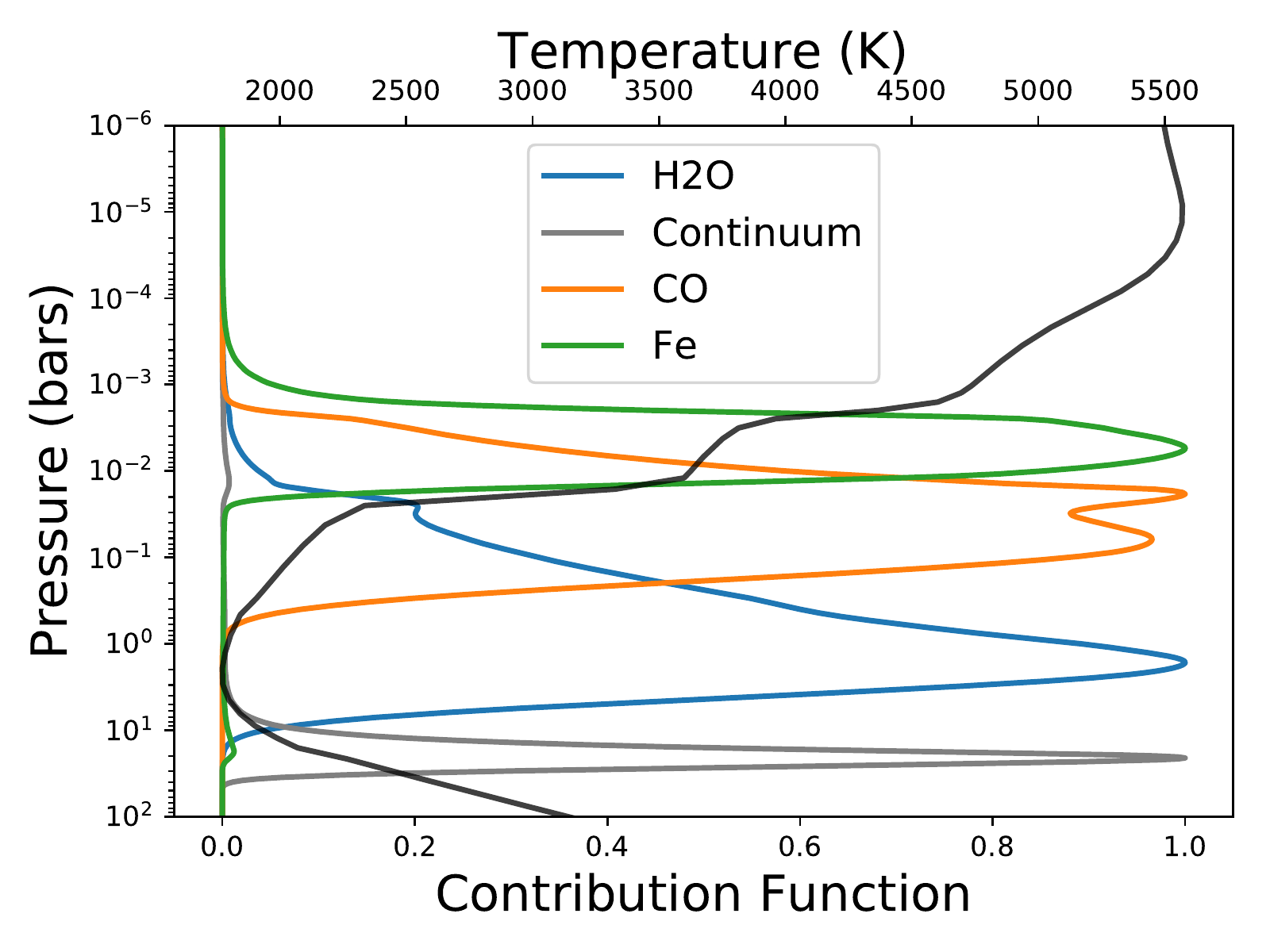} 
	\includegraphics[width=3.5in]{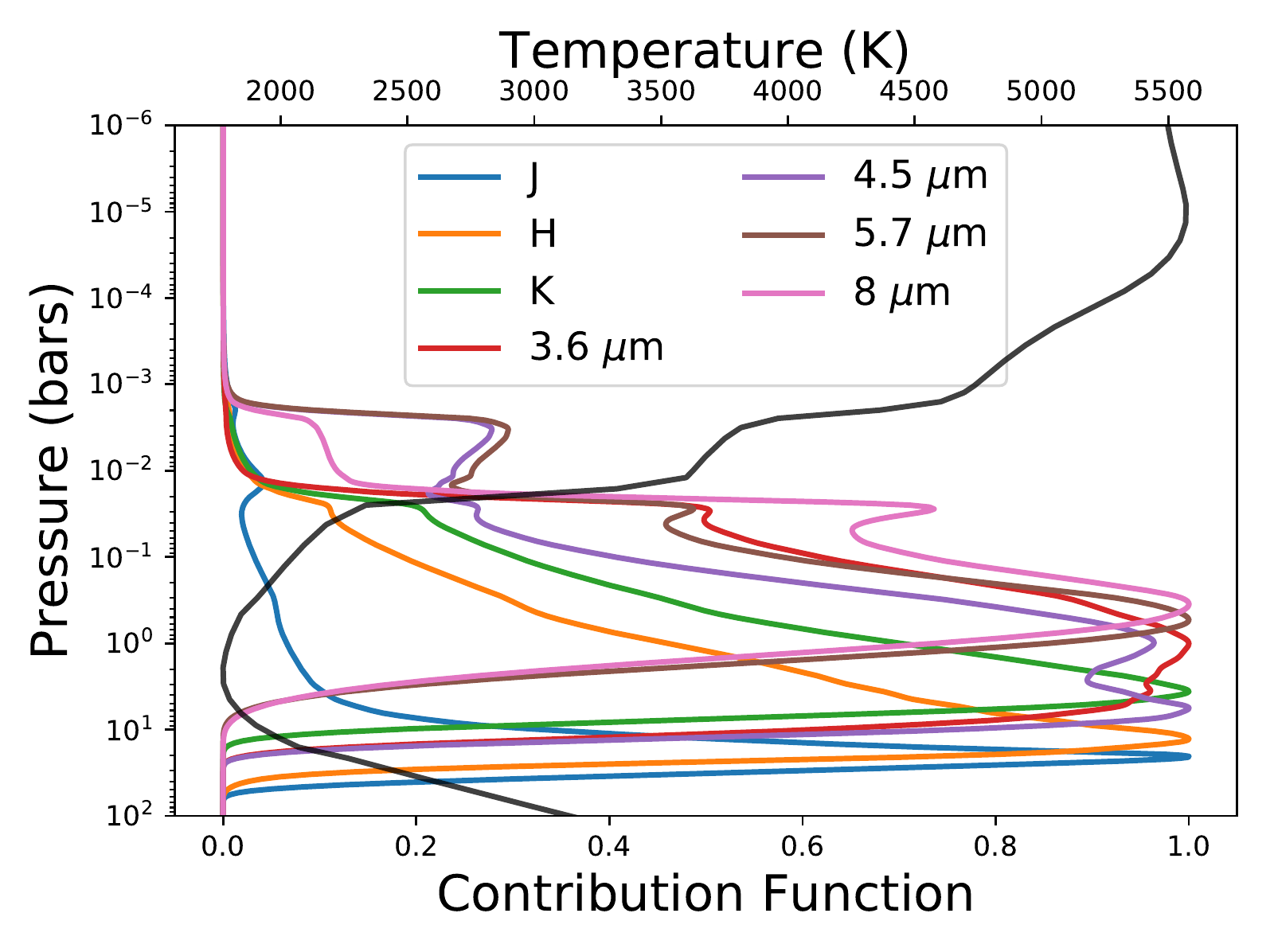}
	\caption{\small Top: Contribution functions of an H$_2$O at 14000~\AA{} (blue), CO at 23048~\AA{} (orange), and Fe at 11886~\AA{} (green) line compared to the contribution function of the continuum at 13000~\AA{} (grey). Plotted behind these against the abscissa on top is the temperature structure of WD-0137B (black). Bottom: same as top, except for the 7 broadband photometric points from \cite{casewell:2015}. \label{fig:wd0137_cont}}
\end{figure}

At a distance of 102.3~pc (333.5~ly) \citep{gaia:2018}, WD 0137-349 is the brightest known white dwarf with a close, detached post-common envelope brown dwarf companion. Low resolution IR spectroscopy and photometry of WD 0137B have previously been best matched with L6-L8 isolated brown dwarf models \citep{maxted:2006,burleigh:2006,casewell:2015}. Higher spectral resolution observations of WD 0137-349B have revealed 42 different emission lines from 11 different atomic species including H I, He I, Na I, K I, Ca I, Ca II, Fe I, Fe II, and Ti I \citep{longstaff:2017}. The equivalent width (EW) of these lines appears to be greatest on the dayside hemisphere of the brown dwarf, indicating these lines may be formed in a temperature inversion caused by irradiation rather than magnetic heating.

To determine whether such a temperature inversion could occur due to irradiation alone and to investigate how the irradiation interferes with the spectral typing of the object, we modeled the dayside atmosphere of WD 0137-349B. The model irradiates WD 0137-349B with a 16,500~K WD with the parameters listed in Table~\ref{table:properties} obtained primarily from \cite{burleigh:2006} and \cite{casewell:2015}\added{ as described in Section~\ref{methods}}.

Figure~\ref{fig:wd0137_spec} shows the resulting spectrum compared to multiband IR photometry \citep{casewell:2015}. The self-consistent model which best fits the observed data\added{, Model 1A}, is cooler than the full heat redistribution model by $\sim$250~K. This results in $T_{\textrm{eff}}$ = 1,840~K. As found in previous comparisons with models \citep{casewell:2015}, it is difficult to simultaneously fit the H, J, and K-band photometry. \added{The $\chi^2$ per data point of this best-fitting self-consistent model\added{, Model 1A,} is only 2.7. This indicates that either our model does not fully describe the data or that the uncertainties on the observations are underestimated. We investigate using atmospheric retrievals to obtain a better match to the data in Section~\ref{sec:wd0137_ret}.} 

An important aspect of the modeled spectrum are the strong emission features. In fact, nearly all the lines detected in emission from the BD by \citep{longstaff:2017} are also found to be in emission in our model. Note, however, that not all spectral features are in emission; while the atomic and CO lines are strongly in emission, H$_2$O is actually in absorption. Similarly, broad spectral lines, like the Na doublet at $\sim$5890~\AA, can exhibit reversed cores, where the wings probe the deeper atmosphere in absorption while the cores show strong emission. This contrasting behavior is quite unique and is a consequence of the concurrent strong external irradiation flux and internal heat flux. Figure~\ref{fig:wd0137_TP} shows the temperature structure of our WD~0137-349B model compared to a similar model without irradiation and to a hot Jupiter of similar integrated irradiation flux (T$_{eq}\sim$1900~K) around a F0V star from \cite{lothringer:2018b}. WD~0137B has a uniquely strong temperature inversion and high internal temperature, creating a temperature minimum in the atmosphere around the location of the IR photosphere, $\sim${1 bar}.

The contribution functions in Figure~\ref{fig:wd0137_cont} demonstrate why H$_2$O is seen in absorption while CO is seen in emission. The contribution function quantifies where in the atmosphere a given wavelength is probing. If the wavelength probes a pressure with a lower temperature than where the continuum probes, a feature will be seen in absorption (i.e., the brightness temperature at wavelengths of that opacity is lower than the continuum). Similarly, if the wavelength probes a pressure with a higher temperature than the continuum, the feature will be seen in emission. CO and strong atomic lines like Fe at 11886~\AA{} probe the hot temperature inversion around 1-100 mbar and will thus be in emission. H$_2$O, however, probes near the temperature minimum of the atmosphere and is cooler than where the continuum probes, which begins to probe the hotter, internal adiabat at $\sim${10 bar}. This phenomenon may provide an alternative explanation (as opposed to TiO cold-trapping) for why some high-mass ultra-hot companions show evidence of H$_2$O absorption, as in KELT-1b \citep{beatty:2017b} and Kepler-13Ab \citep{beatty:2017a}, while their lower-mass counterparts show H$_2$O emission or an isothermal atmosphere \citep{evans:2017,kreidberg:2018,mansfield:2018}.

Also shown in Figure~\ref{fig:wd0137_cont} are the contribution functions for the broadband photometric bands from \cite{casewell:2015}. None of these bands unambiguously probe the temperature inversion and most are centered around 0.1 to 10 bars. IR spectroscopy with, e.g., the James Webb Space Telescope (JWST) will be able to provide a comprehensive understanding of a significant fraction of the lower atmosphere, providing constraints on both the internal temperature and the presence of any temperature inversion.

WD 0137-349B provides an interesting extension to the exploration of temperature inversions in \cite{lothringer:2019}, which compared models  of ultra-hot Jupiters around different host stars. They found that ultra-hot Jupiters around earlier-type host stars will have steeper and stronger temperature inversions due to the increased short-wavelength irradiation flux compared to planets around later-type host stars. \added{\cite{yan:2020} recently showed evidence of this phenomenon with the observation of a $\sim{2,000}$~K temperature inversion in the ultra-hot Jupiter WASP-189b, which orbits an A6 host star \citep{anderson:2018}.} 

At $T_{\textrm{eff}}=16,500$~K, the irradiation from WD 0137-349 is even more weighted towards short wavelengths than any star considered in \cite{lothringer:2019}, where the earliest host star was A0-type with $T_{\textrm{eff}}=10,500$~K. Figure~\ref{fig:wd0137_TP} shows that this increased short-wavelength flux indeed results in an even stronger temperature inversion than a hot Jupiter around a, e.g., F0V type host star. Future observations of irradiated BDs and measurements of their temperature structures will provide essential tests to our understanding of this phenomenon and the overall behavior of temperature inversions in highly irradiated atmospheres.

\subsubsection{Emission Lines Equivalent Widths}

\begin{figure*}[ht!]
	\centering
	\includegraphics[width=6.25in]{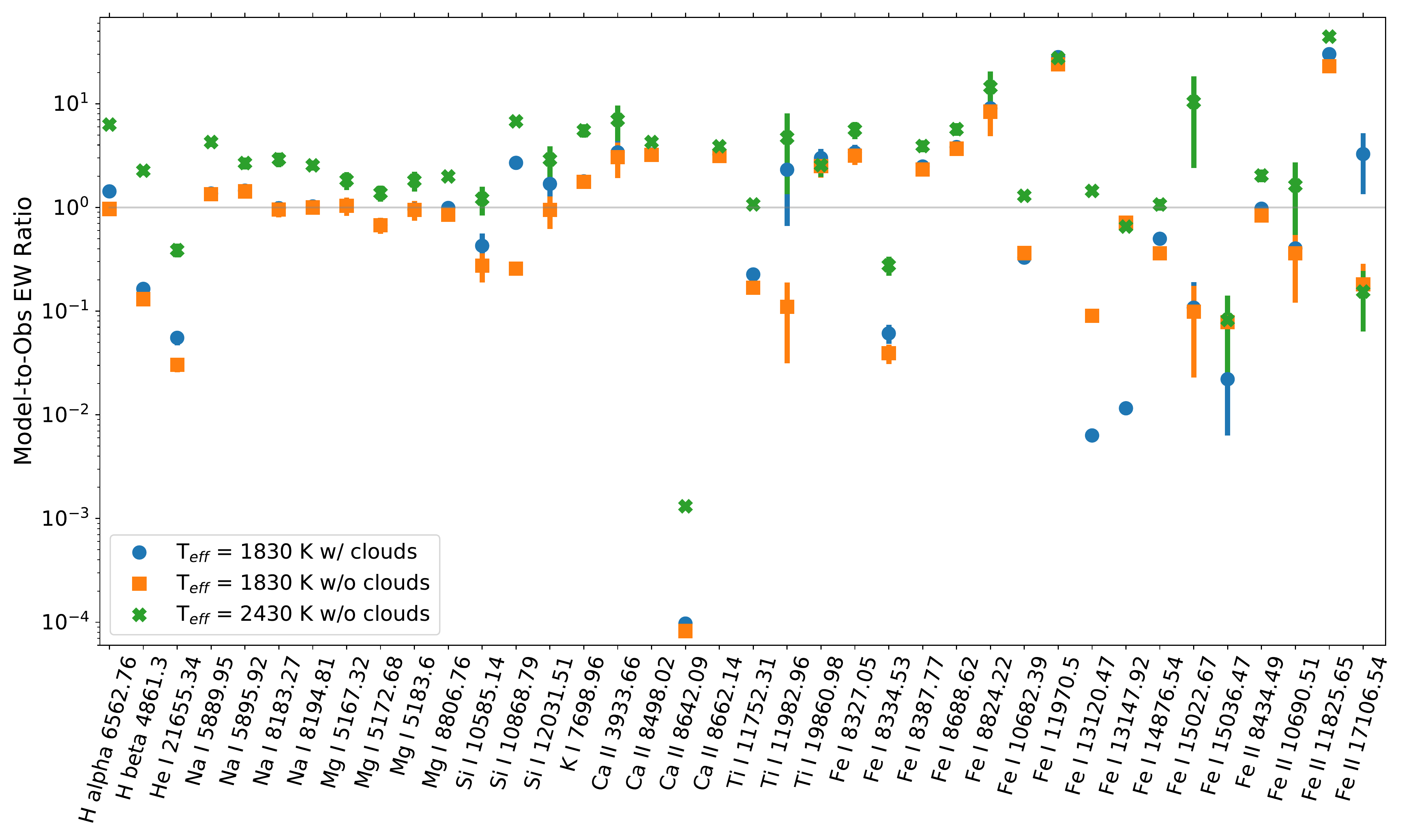} 
	\caption{\small The ratio of the equivalent width of different lines measured in the model to that observed in \cite{longstaff:2017} for three different self-consistent models. A line at 1 is drawn to help identify lines that the model can reproduce well. \label{fig:EWs}}
\end{figure*}

We next investigated whether our self-consistent models could reproduce the equivalent widths (EW) of lines measured in \cite{longstaff:2017}. Figure~\ref{fig:EWs} shows the ratio between the modelled and observed equivalent widths for 38 different lines from 9 different atmospheric species. All lines are detected in emission as in \cite{longstaff:2017}. We did not include the 4 lines that did not have measured EWs in \cite{longstaff:2017}, but did confirm they were in emission. 

Of the measured EWs, the model shown in Figure~\ref{fig:wd0137_spec} can reproduce the H$\alpha$, Na I, Mg I, and K I lines. Results for He I, Si I, Ti I, Fe I \& II are more mixed, with some lines being reproduced, while others are over- or under-estimated. The Ca II appear consistently overestimated.
In general, with the exception of H$\alpha$, the stronger the lines are in the model, the worse they match the observations. This may indicate that our modeled temperature structure matches the atmosphere of WD-0137b where the weaker lines probe (i.e., lower atmosphere), while the temperatures where the strong lines probe (i.e., middle and upper atmosphere) are discrepant. It may also be the fact that some of these lines are exhibiting non-LTE effects due to the very strong irradiation. Lastly, deviations from solar elemental abundances in WD-0137B would also contribute to discrepant modeled EWs. Future modeling and observation could explore these possibilities.

\subsubsection{Retrievals of WD-0137-349B}\label{sec:wd0137_ret}

\begin{figure*}[ht!]
	\centering
	\includegraphics[width=6in]{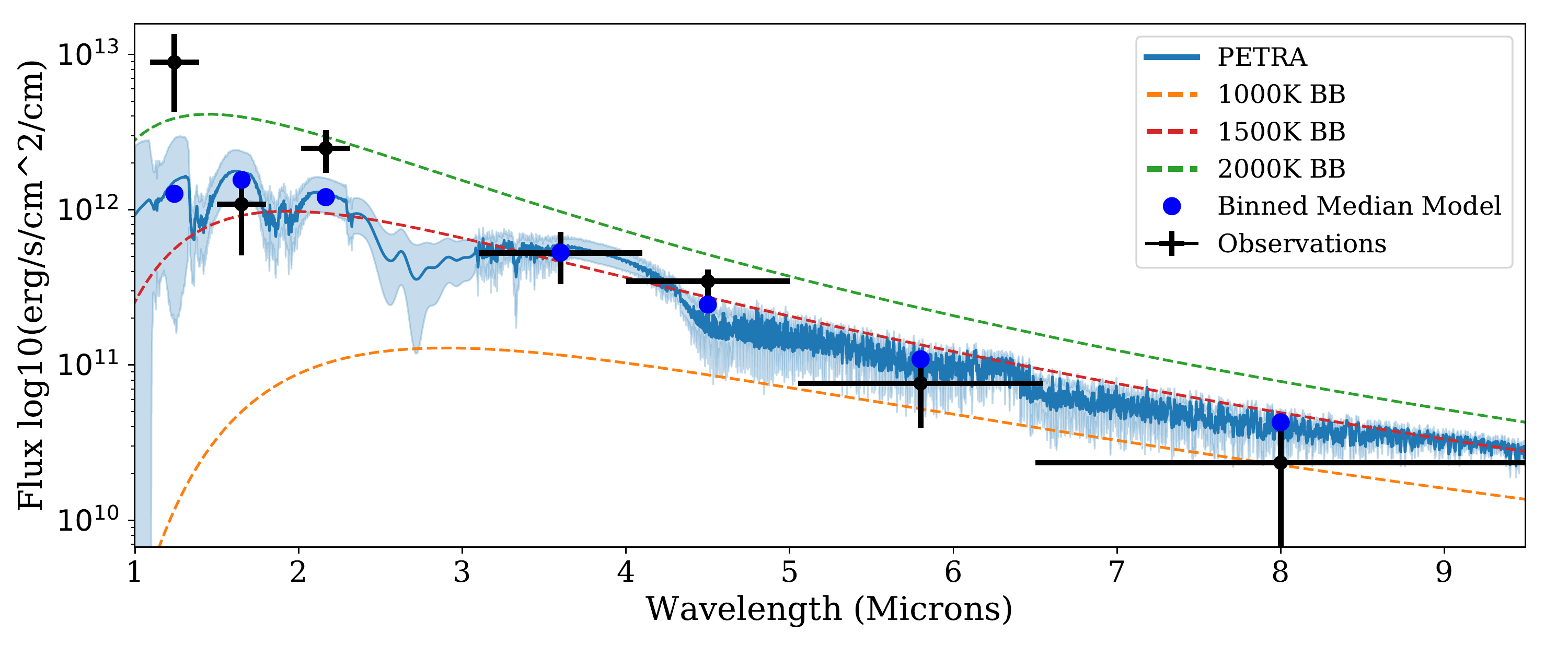} 
	\caption{\small The median retrieved spectrum of WD-0137B and its 1-$\sigma$ uncertainty region compared to the broadband photometry points from \cite{casewell:2015}. Blackbodies of various temperatures are also overplotted as dashed lines. \label{fig:wd0137_petra_spec}}
\end{figure*}

We retrieved the temperature structure of WD-0137B using the broadband photometry from \cite{casewell:2015} with PETRA, for a total of 7 fitted observations and 6 retrieved parameters. Figure~\ref{fig:wd0137_petra_spec} shows the median spectrum from the retrieval and the 1-$\sigma$ uncertainty region compared to the observed photometry. Overall, a better fit was obtained compared to any of the self-consistent models alone, with a best-fit $\chi^2$-per-point of about 1.5\added{, similar to the goodness-of-fit obtained in retrievals of some ultra-hot Jupiters like WASP-121b \citep{evans:2019}}.

\begin{figure}[ht!]
	\centering
	\includegraphics[width=3.5in]{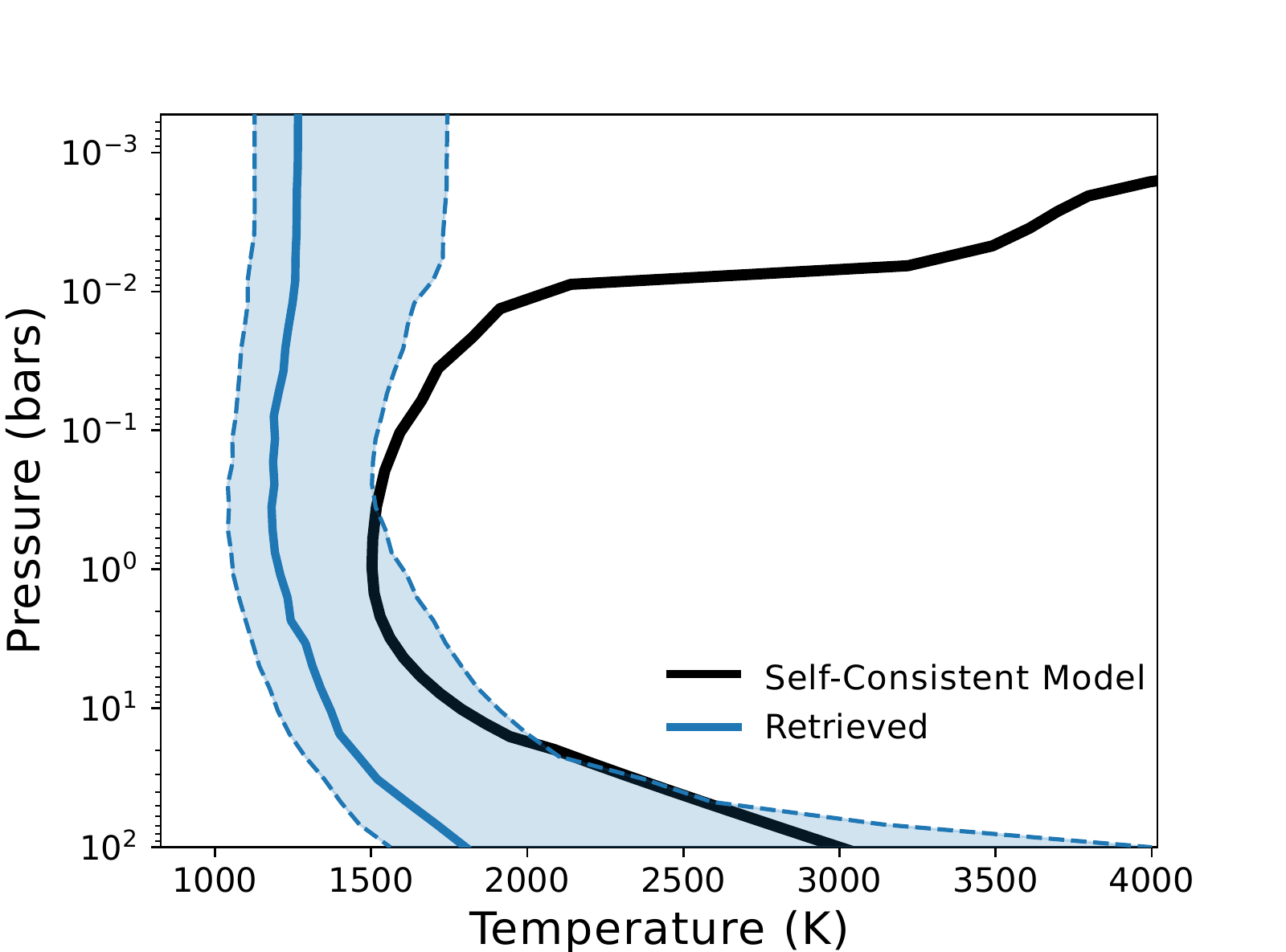} 
	\caption{\small The median retrieved temperature structure of WD-0137B and its 1$\sigma$ uncertainty region (blue) compared to our best-fitting self-consistent model\added{, Model 1A (see Table~\ref{tab:models})}. Evidence for an inversion on WD-0137B comes from atomic emission in spectra not included in this retrieval \citep{longstaff:2017}. \label{fig:wd0137_petra_TP}}
\end{figure}

Figure~\ref{fig:wd0137_petra_TP} shows the retrieved temperature structure compared to \added{Model 1A, }the best-fitting self-consistent model.\added{ The retrieved constraints on the internal and effective temperature are also listed in Table~\ref{tab:models}.} The constraints that PETRA can put on the temperature structure with this dataset are limited, but allow us to measure the overall temperature, which is generally several hundred K cooler than the best-fitting self-consistent model, which itself is $\sim{300}$~K cooler than WD-0137B's equilibrium temperature of 1,990~K.

As described above, Figure~\ref{fig:wd0137_cont} shows where in the atmosphere the observed phototmetry points probe. In general, the photometry probes between 0.1 and 10 bars, implying that any constraints on the temperature structure outside of that pressure range is driven by the parameterization itself, rather than the data. This can explain the absence of an inversion in the retrieved temperature structure: the photometry points simply do not probe where the atmosphere is inverted. Evidence for an inversion on WD-0137B instead comes from atomic emission in spectra not included in this retrieval \citep{longstaff:2017}. Future retrievals can combine the photometry with higher-resolution data to constrain the temperature structure over a larger range in pressure.

From the constraints on the retrieval parameters, we can derive a measurement on the dayside albedo if we assume something about how heat is redistributed across the planet. If we assume full, planet-wide heat redistribution (such that the whole planet is the same temperature), then we calculate an albedo of $0.90^{+0.077}_{-0.15}$. An inefficient heat redistribution, suggested by GCM models \citep{tan:2020,lee:2020}, would seemingly imply an even higher dayside albedo. However, as we discuss in Section~\ref{sec:GCM}, the explanation for the cool temperatures may be due to the fact that we are simply observing the cooler mid-latitude region of the BD around a latitude of 55$^\circ$ because of its $\sim{35}^\circ$ orbital inclination (E. Lee, private communication). Additionally, mineral clouds are likely to form on WD-0137B given its temperature, helping to increase the albedo \citep{lee:2020}. Because the irradiation of these objects is almost entirely in the UV, their albedos may be significantly different than if they were irradiated by a main-sequence star \citep{marley:1999,sudarksy:2000}.

\subsection{EPIC212235321B}

\begin{figure*}[ht!]
	\centering
	\includegraphics[width=3.3in]{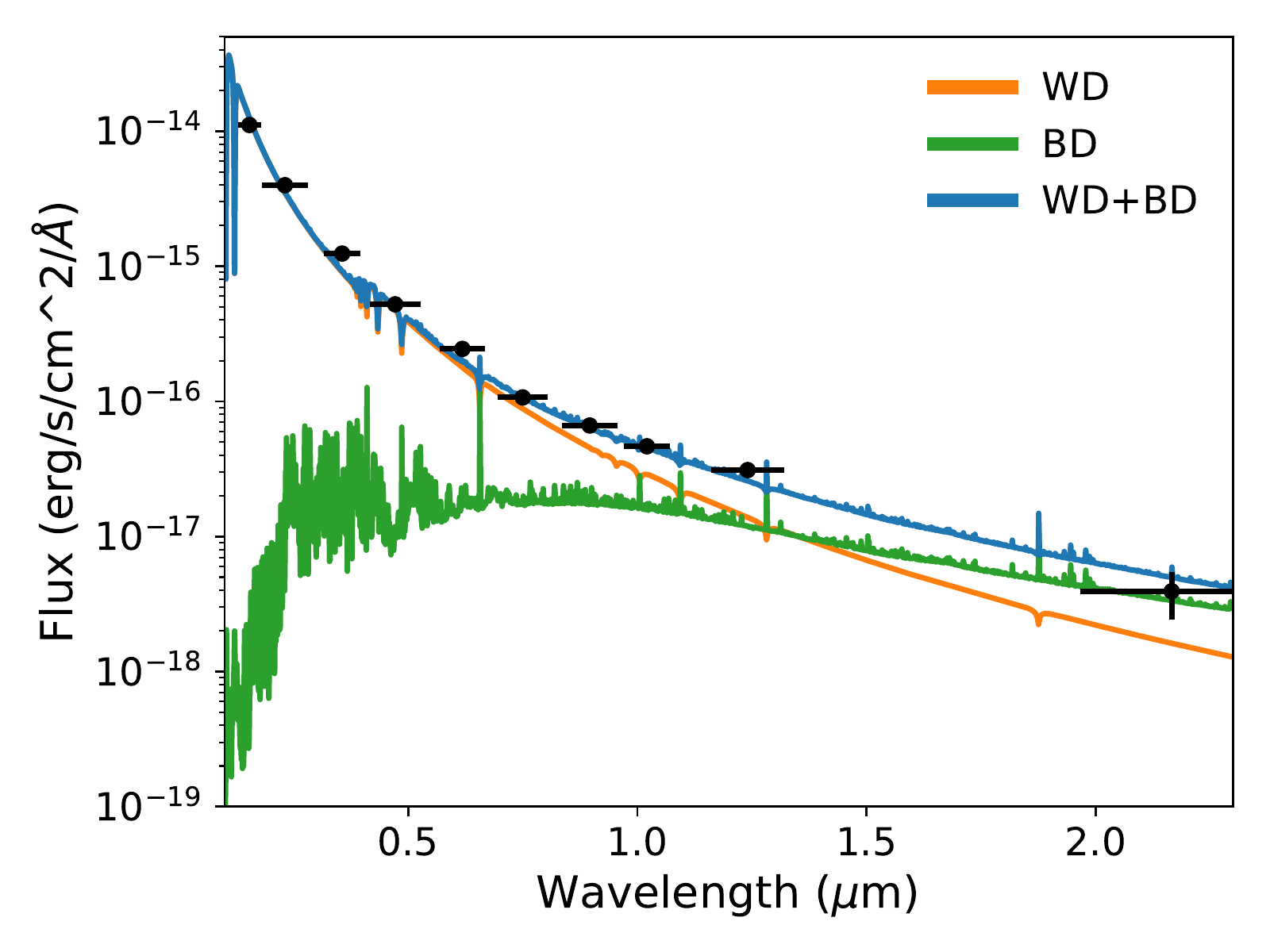} 
	\includegraphics[width=3.3in]{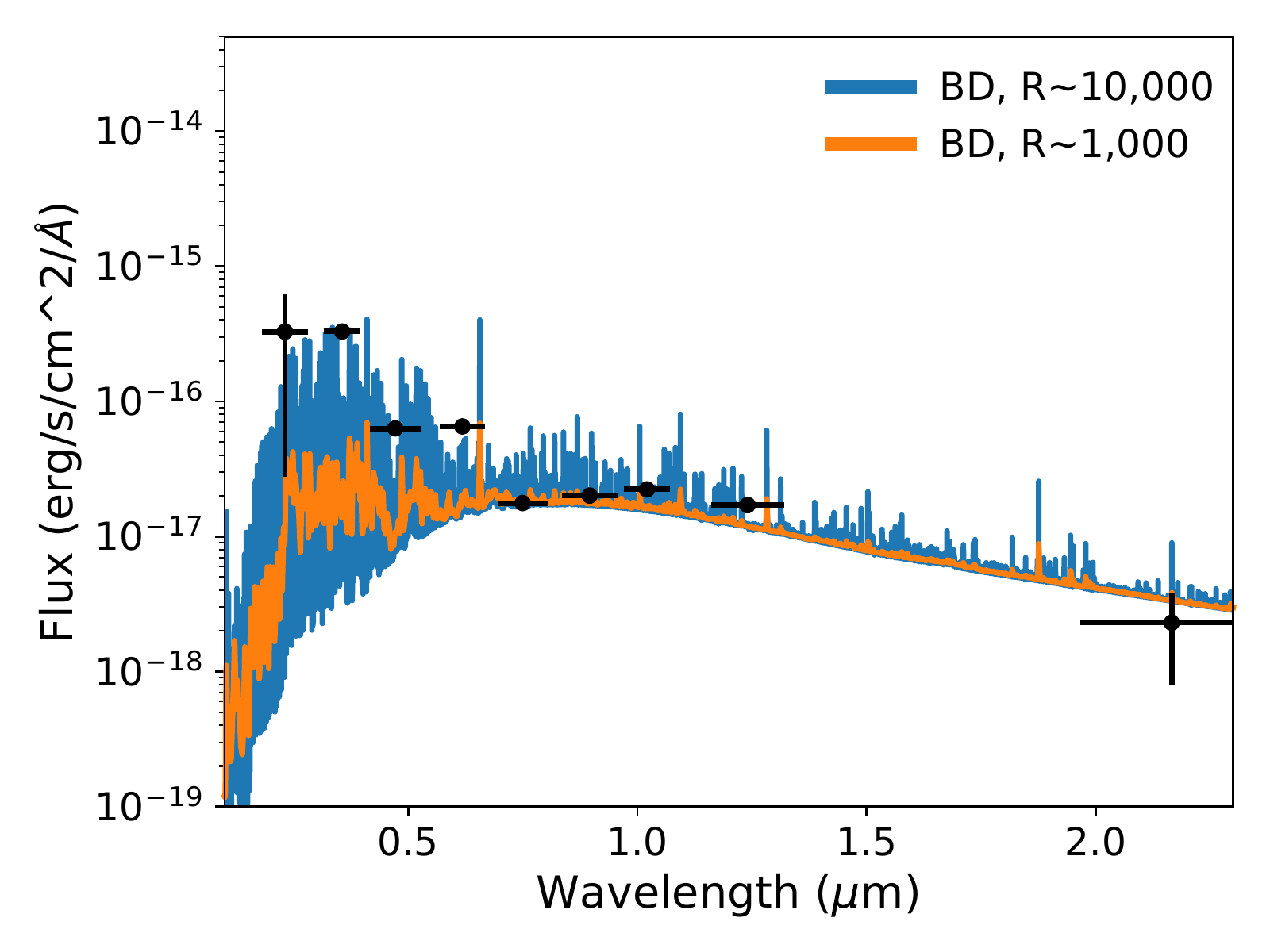}
	\caption{\small Model spectra of the EPIC-2122B system. The left plot shows the WD spectrum (orange), BD spectrum (green), and their combined spectrum (blue) compared to photometry from \cite{casewell:2018}. The right plot shows the BD spectrum at high (R$\sim$10,000 at 1~\microns{}, blue) and lower (R$\sim$1,000 at 1~\microns{}, orange) spectral resolution compared to the same photometry but with the WD contribution subtracted. \label{fig:epic_spec}}
\end{figure*}

The second WD-irradiated BD we consider is EPIC212235321B, which orbits its nearly 25,000~K WD host every 68.2 minutes, making it the shortest-period non-interacting WD-BD system \citep{casewell:2018}. This results in an equilibrium temperature of 3400~K and makes EPIC-2122B hotter than every known hot Jupiter except the $T_{\textrm{eq}}$=4050~K KELT-9b. EPIC-2122B therefore provides an important opportunity to understand the most highly irradiated substellar atmospheres.

William Herschel Telescope/ISIS and VLT/XSHOOTER spectra reveal H$\alpha$ and H$\beta$ emission at the radial velocity of the brown dwarf with evidence for emission in two Mg I and Ca II lines \citep{casewell:2018}. This is suggestive of a temperature inversion and similar to behavior seen in WD-0137-349B and discussed in Section~\ref{sec:wd01347}. Indeed, for atmospheres in this temperature regime, temperature inversions are predicted to be ubiquitous even in planets around main-sequence host stars, caused by the absorption of short-wavelength irradiation by metals like Fe and SiO\added{ and a lack of efficient cooling by molecules like H$_2$O, which begin dissociating at high temperatures} \citep{lothringer:2018b}. Models of even the hottest isolated brown dwarfs ($\sim$M8) underestimate the observed photometry so irradiated models are necessary to interpret current observations \citep{casewell:2018}.

Our model used the parameters listed in Table~\ref{table:properties} taken from \cite{casewell:2018}. We assumed solar metallicity and full heat redistribution for our fiducial model. We calculated models for T$_{int}$ = 1000 and 2000~K. We use an orbital separation of 0.44~$\rsun$ and a distance of 386.8 pc (1262 ly) from Earth \citep{gaia:2018}. 

Figure~\ref{fig:epic_spec} shows the model WD and BD spectra compared to observed photometry from \cite{casewell:2018}. The photometry for the BD alone was calculated by subtracting a model for the WD component from the total flux. Our modeled spectrum fits significantly better than the best-fit isolated model from \cite{casewell:2018}, undoubtedly because of the self-consistent treatment of the intense irradiation. With that said, the computed $\chi^2$ per data point for all fits to the EPIC2122B observations are quite poor (see Table~\ref{tab:models}). The full heat redistribution model ($f=0.25$) with T$_\textrm{int}$=2,000~K\added{, Model 2B,} fits the data the best, but has a $\chi^2$ per data point of 26.1. The uncertainties on the data a very small and an underestimation of noise is possible, but an incorrect assumption in the model (e.g., 1D atmosphere, solar metallicity, chemical equilibrium) may simply be amplified by the precise measurements.

Besides atomic emission lines, the IR wavelengths are relatively featureless because of strong H$^{-}$ continuous opacity and the thermal dissociation of molecules, much like ultra-hot Jupiters (see Section~\ref{sec:HJs}). The NUV and optical photometry is still underestimated, but this is likely from residuals to our WD model fit in that region, which overwhelms any contribution from the BD. Our models also reproduce the H$\alpha$, H$\beta$, Mg I, and Ca II emission lines seen in the XSHOOTER and ISIS spectra. Our synthetic spectra suggest a rich forest of other atomic emission lines that could be observable with high-resolution and high-SNR spectroscopy.

\begin{figure}[ht!]
	\centering
	\includegraphics[width=3.4in]{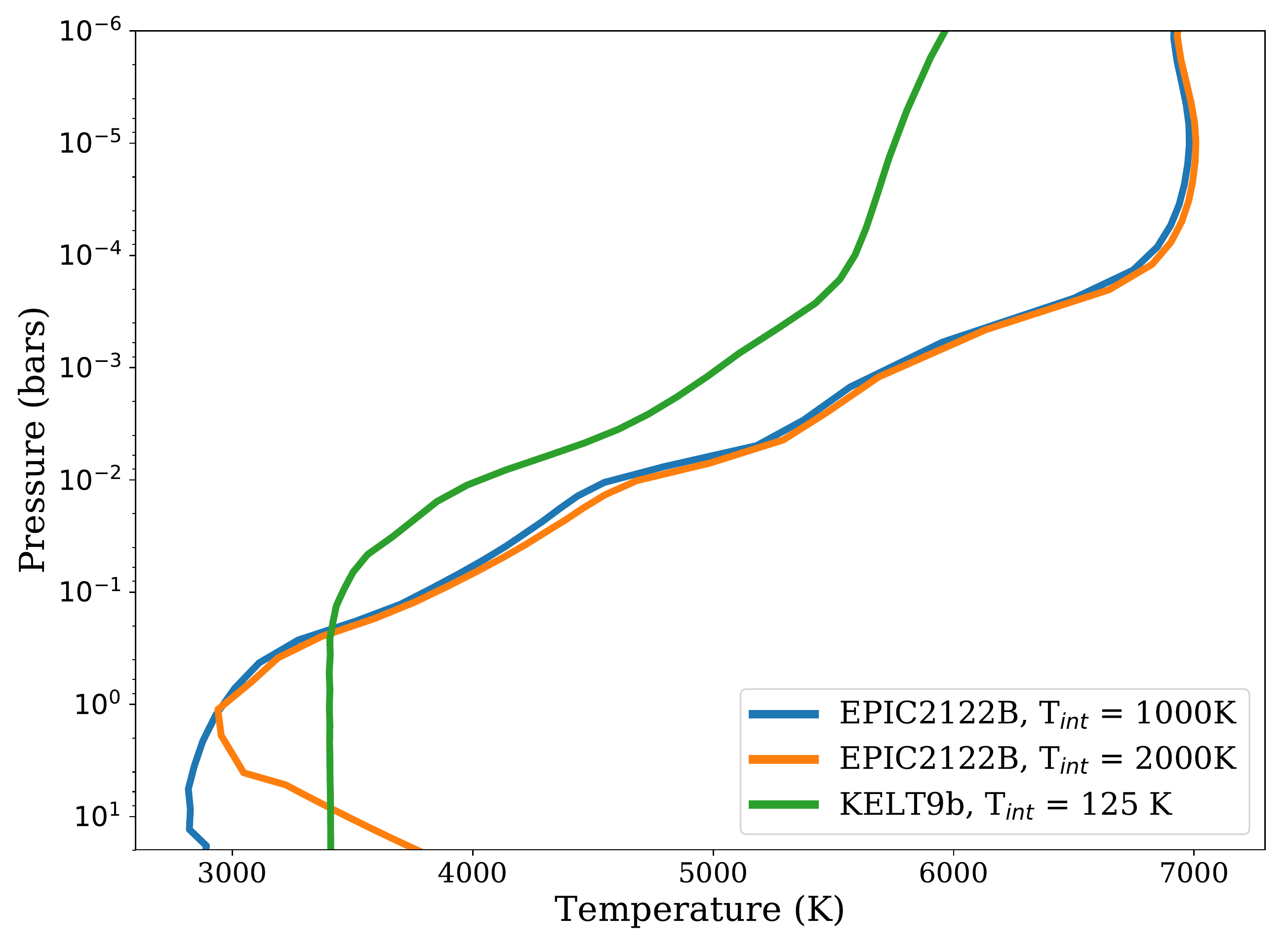} 
	\caption{\small Temperature structures from models of EPIC-2122B assuming T$_{\textrm{int}}$= 1000 (blue) and 2000~K (orange) compared to a model of KELT-9b (T$_{\textrm{int}}$= 125 K). Note that internal temperatures from hot Jupiters may be up to several hundred Kelvin hotter \citep{thorngren:2019}. Also note KELT-9b and EPIC2122B will have very different photospheric pressures because of their different surface gravity. \label{fig:epic_struct}}
\end{figure}

Figure~\ref{fig:epic_struct} shows the temperature structures from our model atmospheres of EPIC-2122b, assuming T$_{int}$ = 1000 and 2000~K compared to KELT-9b. Surprisingly, while EPIC-2122b has a lower equilibrium and effective temperature, the intense short-wavelength irradiation from the WD host is able to generate a temperature inversion with higher maximum temperatures than KELT-9b. Note though that KELT-9b and EPIC-2122b have very different surface gravity (\logg = 3.2 versus 5.2, respectively), so the photospheric pressures will also be very different.

\subsubsection{Retrievals of EPIC-2122B}

In addition to computing self-consistent models, we also ran a retrieval on the observed broadband photometry from \cite{casewell:2018}. Like our retrievals of WD-0137B, we only retrieved for the temperature structure given the low number of points while fixing the model to solar metallicity and assuming chemical equilibrium. We neglected points shortward of I-band because they are dominated by the residuals from the subtraction of the WD component, as mentioned above.

\begin{figure*}[ht!]
	\centering
	\includegraphics[width=6in]{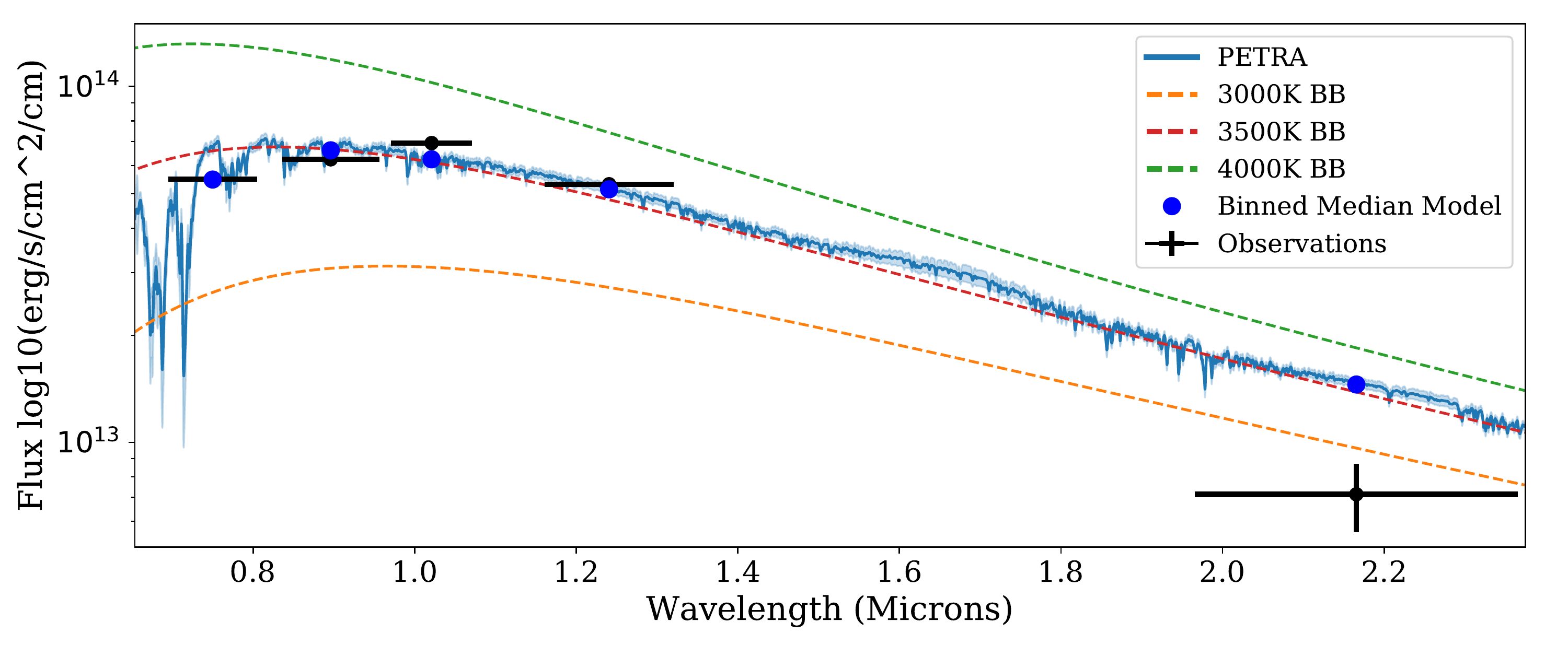} 
	\caption{\small The median retrieved spectrum of EPIC 2122B and its 1-$\sigma$ uncertainty region compared to the photometry from \cite{casewell:2018}. Blackbodies of various temperatures are also overplotted as dashed lines. \label{fig:epic_petra_spec}}
\end{figure*}

Figure~\ref{fig:epic_petra_spec} shows the median retrieved spectrum versus the observations. \added{With a $\chi^2$ per data point of 18.6}, the best-fit from the retrieval is somewhat better than any of the self-consistent models to which we compared. The observations closely follow a 3,500~K blackbody, except for the low K-band point. The I-band point is somewhat below a 3,500~K brightness temperature and is well fit by absorption from TiO\added{ at a few bars}, more comparable to an M-dwarf than an L-dwarf. Like the spectra from the self-consistent models above, the retrieved spectrum is relatively featureless except for atomic emission lines due to the high temperature resulting in significant H$^{-}$ continuous opacity, explaining the blackbody-like behavior of the spectrum.

\begin{figure}[ht!]
	\centering
	\includegraphics[width=3.5in]{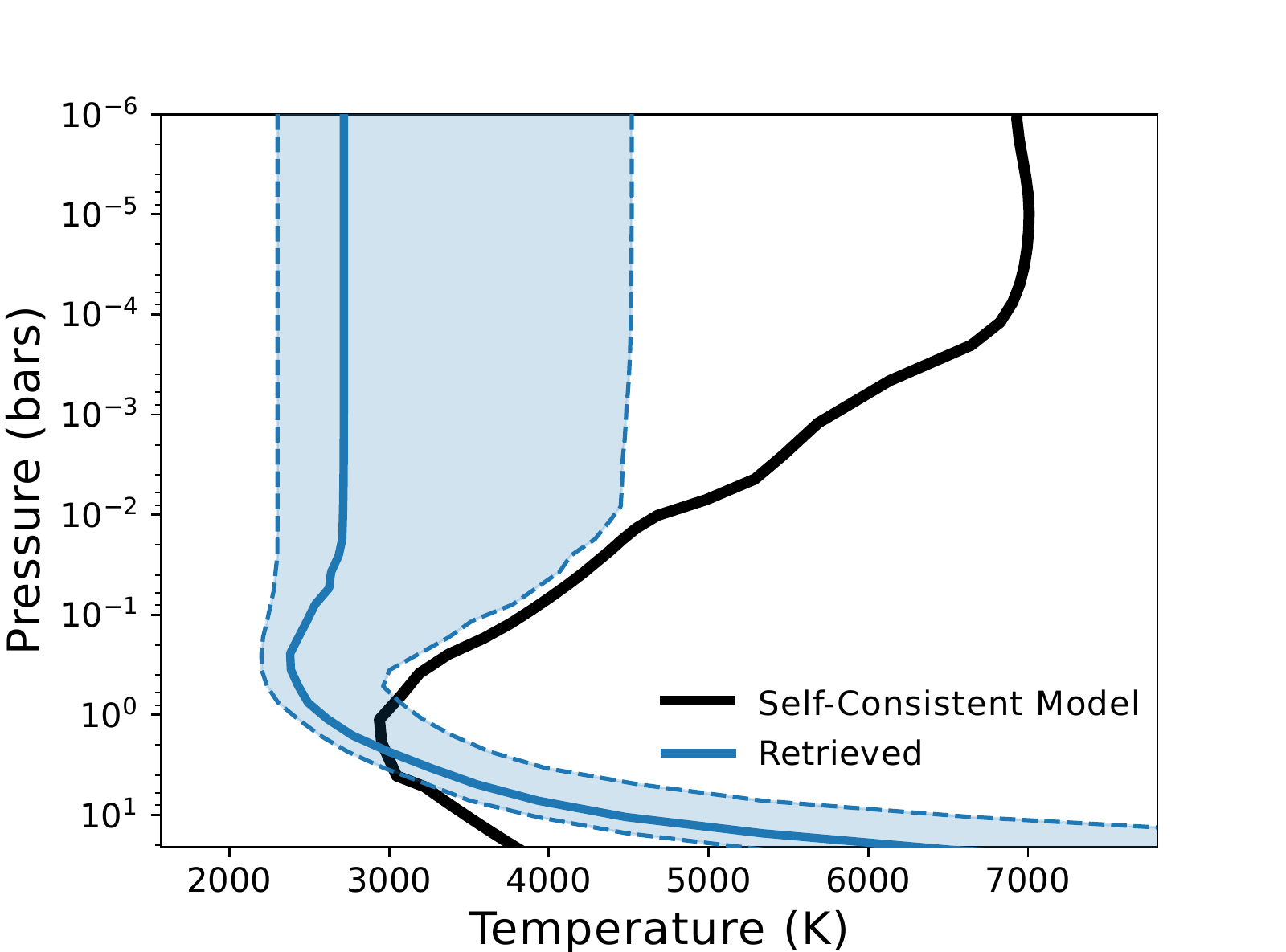} 
	\caption{\small The median retrieved temperature structure of EPIC2122B and its 1$\sigma$ uncertainty region (blue) compared to the fiducial, full-heat-redistribution self-consistent model with T$_{\textrm{int}}=2000$~K. \label{fig:epic_petra_TP}}
\end{figure}

Figure~\ref{fig:epic_petra_TP} shows the retrieved temperature structure. Though constraints are limited, the data seem to suggested an even hotter interior temperature than the model with T$_{\textrm{int}}=2000$~K shown in Figure~\ref{fig:epic_petra_TP}. Constraints on the internal and effective temperatures are listed in Table~\ref{tab:models}. Like WD-0137B, the broadband photometry do not provide much information on the temperature inversion, but some of the retrieved temperature profiles are consistent with an inversion beginning around 0.5 bar.

\section{Discussion}\label{discuss}

\subsection{Comparison to Hot and Ultra-Hot Jupiters}\label{sec:HJs}

While WD-0137B has an equilibrium temperature similar to many hot Jupiters like HD 189733b and HD 209458b, our models predict quite different behavior in the temperature structures. For hot Jupiters warmer than about 1,700-1,800~K, 1D chemical equilibrium models predict temperature inversions caused by the absorption of irradiation by TiO and VO \citep{hubeny:2003,fortney:2008}. 

\begin{figure}[h!]
	\centering
	\includegraphics[width=3.5in]{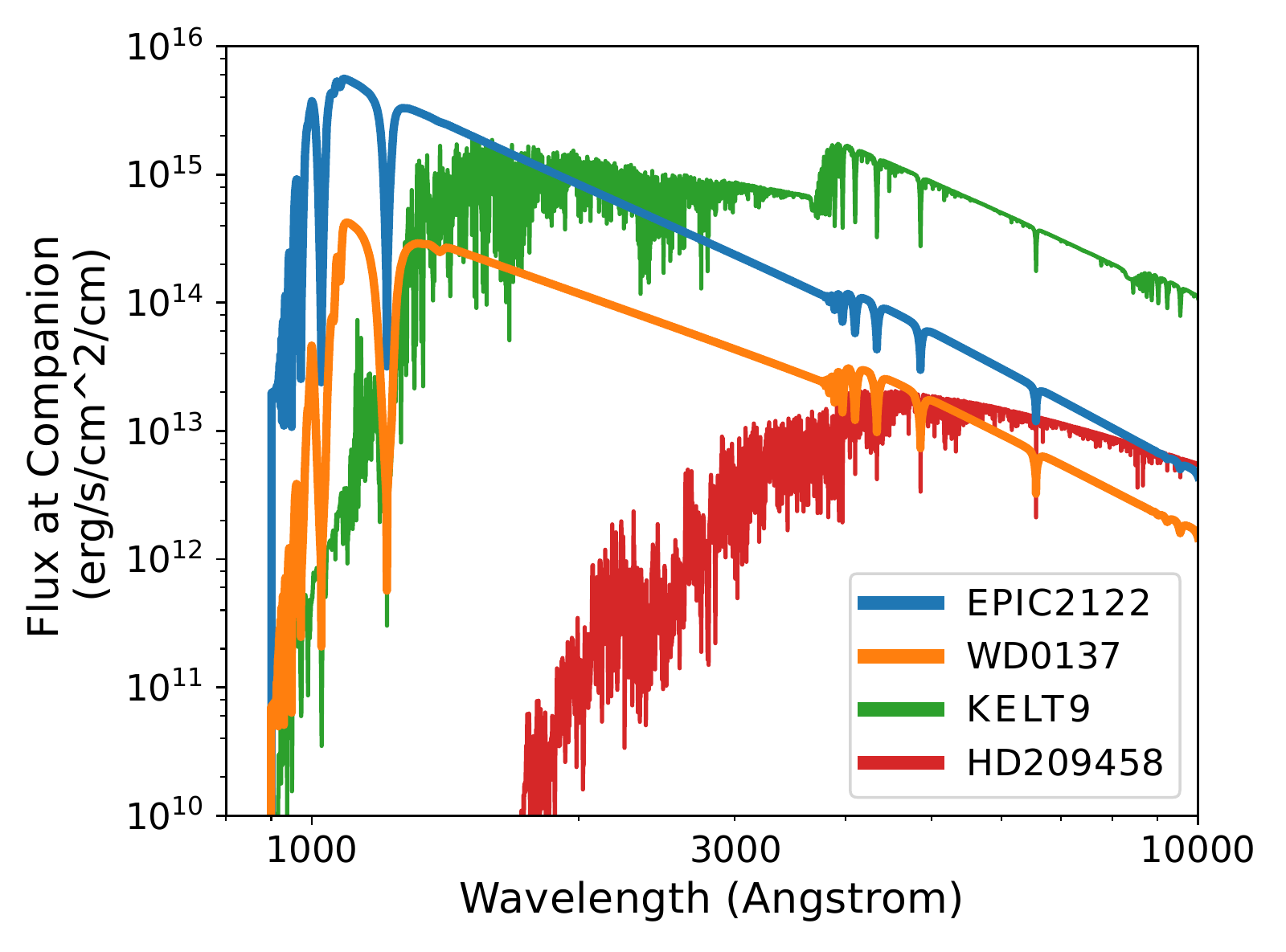} 
	\caption{\small Irradiation flux at the location of the companion for 4 systems: EPIC-2122 (blue), WD-0137-349 (orange), KELT-9 (green), and HD~209458b (red). \label{fig:stars}}
\end{figure}

TiO and VO play an especially important role in the energy balance of the atmosphere because their opacity is significant precisely where much of the irradiation is output from main-sequence host stars, namely around 5000-8000 \AA. For BDs irradiated by WDs, however, the irradiation is much more heavily weighted towards the UV. This is demonstrated in Figure~\ref{fig:stars}, which shows the irradiation as a function of wavelength at the location of WD-0137B, EPIC2122B, KELT-9b, and HD209458b.

In WD-0137B, a temperature inversion does not form because of TiO and VO, molecules which will absorb little of the total incoming flux and may even be condensed in parts of the atmosphere. Rather, it is the absorption of short-wavelength irradiation by gaseous metals and molecules like SiO. Thus, WD-0137B behaves much like the hottest ultra-hot Jupiters like KELT-9b, despite experiencing a total integrated flux similar in magnitude to a Jupiter in the hot-to-ultra-hot transition.

Figures~\ref{fig:epic_spec} and \ref{fig:epic_struct} demonstrate how similar EPIC2122b is to ultra-hot Jupiters. This is also true for other extremely irradiated BDs in WD systems, like WD1202-024B \citep{rappaport:2017}. Such systems will all exhibit spectra with weak or absent molecular features due to a combination of H$^{-}$ raising the photosphere and molecular dissociation at high temperatures \citep{arcangeli:2018,lothringer:2018b,parmentier:2018,kitzmann:2018}. CO, the strongest molecule in nature, can usually remain intact about an order-of-magnitude lower in pressure compared to other absorbers like H$_2$O and TiO. Observations of the most highly irradiated planets and brown dwarfs can search for CO to help constrain the temperature and/or chemical abundances. Additionally, our models predict rich forests of atomic lines are likely to be found in irradiated BDs. This is supported by detections of various gaseous metals in the atmospheres of both ultra-hot Jupiters and irradiated BDs \citep[e.g.,][]{longstaff:2017,casewell:2018,hoeijmakers:2019,lothringer:2020b}.

\subsection{Irradiated BDs in Color-Magnitude Space}

Brown dwarf companions to white dwarfs have historically been classified according to the spectral types defined for isolated objects \citep{kirkpatrick:1999,burgasser:2001,kirkpatrick:2012}. The shortcomings of this scheme are illustrated with EPIC2122B, for which the near-IR photometry suggested types earlier than M8, implying that the spectral energy distribution was affected by the irradiation \citep{casewell:2018}. Figures~\ref{fig:wd0137_TP} and \ref{fig:epic_struct} have shown how the temperature structure of irradiated BDs will look very different from isolated objects. We therefore investigate how irradiation will change an objects behavior throughout color-magnitude space. Such studies have previously been carried out in the context of hot Jupiters \citep{triaud:2014,triaud:2014b,manjavacas:2018,melville:2020,dransfield:2020}.

For this exercise, we chose the parameters of the full-heat redistribution ($f$ = 0.25) WD-137B model as the fiducial model. We then changed the internal heat flux, through T$_{\textrm{int}}$, and the irradiation flux by adjusting the orbital separation of the system. We quantify the reduction in irradiation flux from Eq.~\ref{eq1}. These parameters are listed in Table~\ref{table:properties_CMD} for the models of WD-0137B we computed. For each of these sets of parameters, cloud-free and  cloudy models were calculated, as described in Section~\ref{methods}.

\begin{figure*}[ht!]
	\centering

	\gridline{\fig{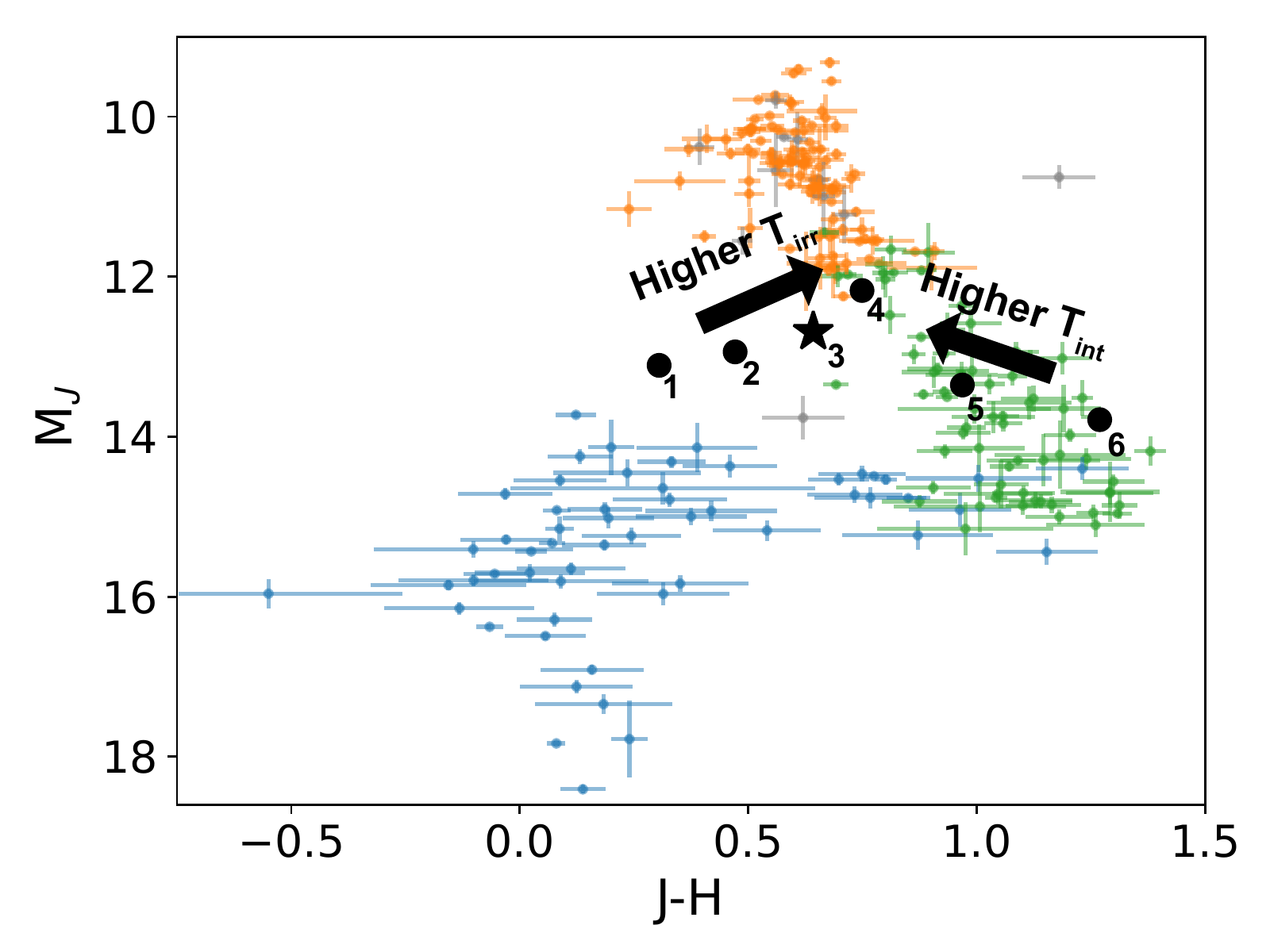}{0.5\textwidth}{\large a) Cloud-Free}
	\fig{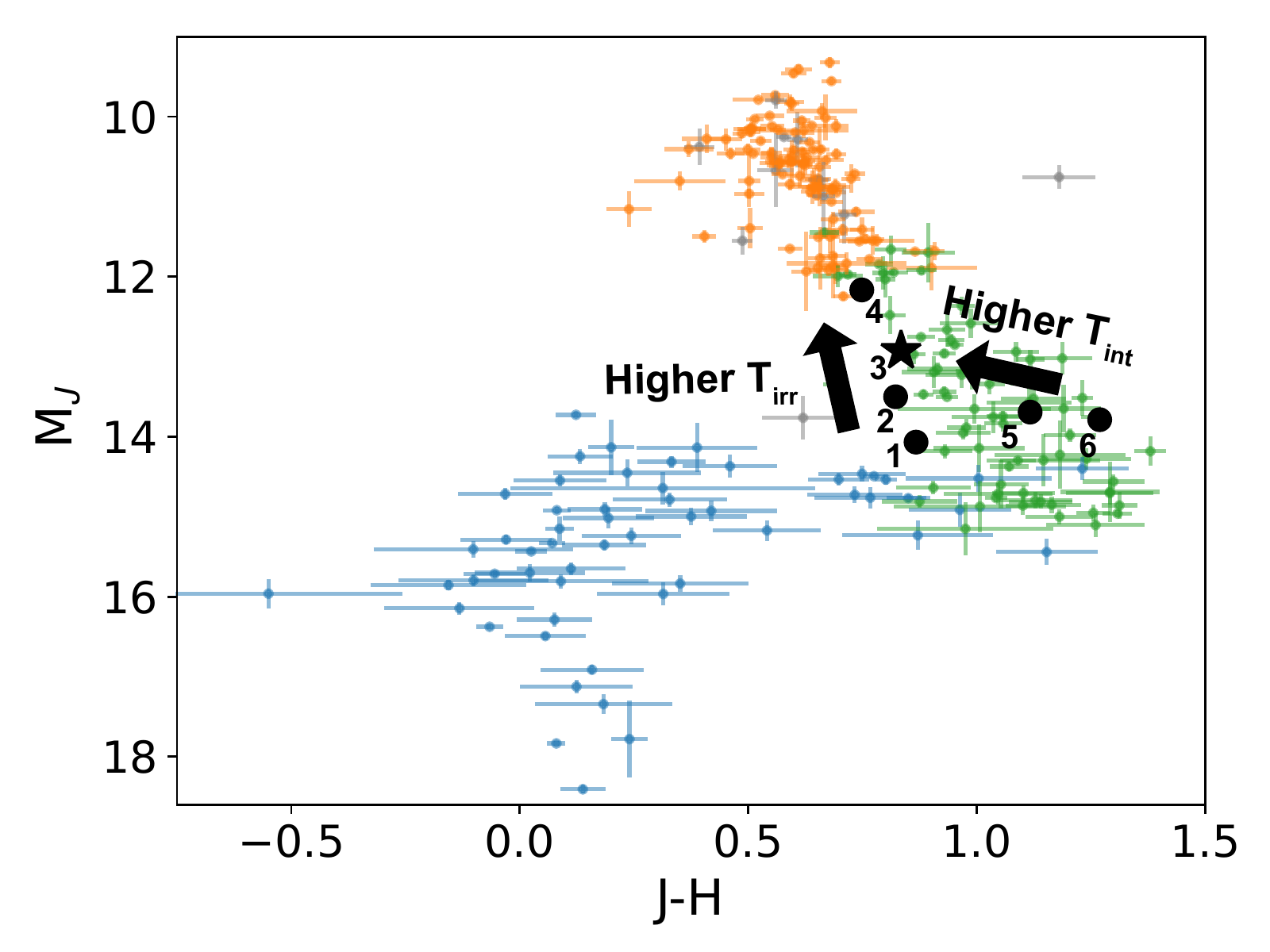}{0.5\textwidth}{\large b) Cloudy}}

	\caption{\small The behavior of different parameters on color-magnitude diagrams. Black points vary the irradiation, represented by T$_\textrm{irr}$, and grey points vary the internal temperature, T$_\textrm{int}$ for cloud-free (left) and cloudy (right) atmospheres. The yellow star represents the fiducial model with T$_\textrm{int}$ = 1400~K and T$_\textrm{irr}$ = 2000~K. Underneath are plotted are the measured colors of M-dwarfs (orange), L-dwarfs (green), and T-dwarfs (blue) from \cite{dupuy:2012}. Magnitudes use or were converted to the 2MASS J, H, K$_{\rm s}$ filter system. \label{fig:wd0137_cmd}}
\end{figure*}

Figure~\ref{fig:wd0137_cmd} shows the location of each of these models in absolute J magnitude ($M_\textrm{J}$) versus J-H color space compared to the location of isolated brown dwarfs from \cite{dupuy:2012}. In both the cloudy and cloud-free models, Model 1, which has no irradiation ($T_{\textrm{irr}}$=0), illustrates the location of WD-0137B if it indeed were isolated. As $T_{\textrm{irr}}$ is increased, WD-0137B gets redder if it is cloud-free and bluer if it is cloudy. As $T_{\textrm{int}}$ is increased, WD-0137B becomes bluer in both the cloudy and cloud-free cases.

This complex behavior can mean that a highly irradiated brown dwarf can be misclassified as an object of a spectral type that does not represent the true T$_{\textrm{int}}$, T$_{\textrm{irr}}$, or T$_{\textrm{eff}}$. For example, if WD-0137B is cloud-free, it could be misclassified as a cloudy object with higher T$_{\textrm{int}}$. Similarly, a cloudy WD-0137B could be misclassified as a cloudless object with low T$_{\textrm{int}}$ and T$_{\textrm{irr}}$. The implication of this fact is that these irradiated brown dwarfs cannot be straightforwardly characterized by their colors alone, especially using schemes for isolated objects. Dedicated modeling, both self-consistent and retrieval, are likely necessary to classify these complex objects. 

 \begin{table}[t] 
	\centering  
	\caption{WD-0137B CMD Model Parameters}
	\label{table:properties_CMD}
	\hspace{-55pt}
	\begin{tabular}{c|ccc}
		
		\hline
		Model & T$_{\textrm{irr,BD}}$ (K) &  T$_{\textrm{int,BD}}$ (K) & T$_{\textrm{eff,BD}}$ (K) \\
		\hline 
		\hline
		1  & 0 & 1,400 & 1,400 \\ 
		\hline
		2  & 1,660 & 1,400 & 1,840 \\ 
		\hline
		3\tablenotemark{\footnotesize a} & 1,990 & 1,400 & 2,100 \\ 
		\hline 
		4  & 2,360 & 1,400 & 2,430 \\ 
		\hline 
		5  & 1,990 & 1,000 & 2,020 \\ 
		\hline		
		6  & 1,990 & 400 & 1,991 \\
		\hline 
		
	\end{tabular}
	\tablenotetext{\footnotesize a}{\footnotesize Fiducial model}
\end{table} 

\subsection{Comparison to GCM Results}\label{sec:GCM}
Lastly, we compare our results to those of two recent studies that investigated the circulation in BDs orbiting WDs \citep{tan:2020,lee:2020}. The combination of intense irradiation and extremely short period put these objects in an interesting and relatively unexplored dynamical regime. Additionally, all hot Jupiter phase curves to date probe transiting planets, with the exception of $\upsilon$ Andromedae b \citep{crossfield:2010}. This means that nearly all of our knowledge of hot Jupiter circulation comes from observing the equatorial regions. The non-transiting nature of many BDs in WD systems provide us an opportunity to probe the mid-latitude circulation of sub-stellar atmospheres.

As mentioned above, WD-0137-349B exhibits a cooler atmosphere than predicted by our 1D dayside atmosphere models. This was also shown in \cite{lee:2020} who were able to match the shape of WD-0137-349B's phase curves with their GCM model, but over-predicted the absolute flux by up to a factor of 3 in some bands. While the assumptions and simplicity of the retrieval model combined with the limited data set prevent an unambiguous interpretation of the temperature structure, the explanation for the low retrieved temperature is potentially related to the $\sim${35}$^\circ$ orbital inclination of WD-0137B. This means that rather than probing the hotter equatorial region, observations probe the mid-latitude region around 55-35$^\circ$. Indeed, our retrieved temperature structure is similar in temperature to mid-latitude regions in GCMs. \added{A similar effect may also be responsible for the low temperatures retrieved for EPIC2122B at pressures below a bar, though the retrieved temperature for the lower atmosphere is hotter than our model predictions.}

Compared to a hot Jupiter, BDs in short-period WD systems will have a much narrower equatorial jet due to the high gravity and fast rotation leading to a small Rossby deformation radius. These effects compress the equatorial jet and result in a cooler mid-latitude temperature. 

An important next step in studying these irradiated BDs will be the combination of GCMs with radiative schemes to understand the effect their extreme UV irradiation has on the circulation. Here, we've shown that the UV irradiation can play an important role in the behavior of the temperature structure.

\section{Conclusion}\label{conclusion}

We have presented self-consistent and retrieved models of two brown dwarfs irradiated by white dwarfs, WD-0137-349B and EPIC-212235321B, that take into account the effect of the intense UV irradiation on the atmosphere. Overall, our models and synthetic spectra match previously observed photometry and can help explain the detections of atomic emission coming from both WD-0137B and EPIC-2122B \citep{longstaff:2017,casewell:2018}.

These objects are analogous to but significantly different from similarly irradiated hot Jupiters. Our models indicate that objects like WD-0137B can simultaneously exhibit absorption and emission spectral features because of the potentially high internal temperature and the irradiation-induced temperature inversion. Future high-precision spectroscopy from ground-based and space-based facilities (like JWST) could test for the existence of such behavior while further characterizing the temperature structure, composition, and circulation in these atmospheres.

Using the existing photometry, we have also retrieved the temperature structure for WD-0137B and EPIC-2122B with PETRA. While the constraints are limited by the low number of data points, both temperature structures appear consistent with our models, but are slightly cooler, potentially due to the fact that we are observing the mid-latitude region of the BDs.

While short-period BDs orbiting WDs are rare, they provide a valuable laboratory to explore the physics of both substellar and irradiated atmospheres. The unique presence of strong internal heat and intense irradiation places these objects in a unique corner of parameter space compared to isolated BDs and hot Jupiters. The fact that the irradiation is almost entirely at UV wavelengths further tests our understanding. Observation and modeling of these extreme systems will teach us about the processes at work in all types of atmospheres.

	\acknowledgments
	We thank the anonymous reviewer for a helpful report. We would also like to thank Jonathan Fortney for making sure the two authors met at Exoclimes V. SLC acknowledges the support of an STFC Ernest Rutherford Fellowship. This research has made use of NASA's Astrophysics Data System Bibliographic Services.

	\bibliographystyle{aasjournal}

\end{document}